\documentclass{article}
\usepackage{iclr2026/iclr2026_conference,times}

\usepackage{hyperref}
\usepackage{url}
\usepackage{microtype}
\usepackage{graphicx}
\usepackage{subcaption}
\usepackage{booktabs}
\usepackage{enumitem}
\usepackage{multirow} 
\usepackage{amssymb}
\usepackage{amsmath}
\usepackage{mathtools}
\usepackage{amsthm}
\usepackage{placeins}

\iclrfinalcopy

\newcommand{\cmark}{\checkmark}
\newcommand{\xmark}{\texttimes}

\theoremstyle{plain}
\newtheorem{corollary}{Corollary}[section]
\theoremstyle{definition}
\newtheorem{definition}{Definition}[section]
\theoremstyle{remark}
\newtheorem{claim}{Claim}[section]

\title{Validating Interpretability in siRNA Efficacy Prediction: A Perturbation-Based, Dataset-Aware Protocol}

\author{Zahra Khodagholi\thanks{Corresponding author: \texttt{Zahra.Khodagholi@ucf.edu}} \\
University of Central Florida \\
\texttt{Zahra.Khodagholi@ucf.edu} \\
\And
Niloofar Yousefi \\
University of Central Florida \\
\texttt{Niloofar.Yousefi@ucf.edu}
}

\begin{document}

\maketitle

\begin{abstract}
Saliency maps are increasingly used as \emph{design guidance} in siRNA efficacy prediction, yet attribution methods are rarely validated before motivating sequence edits. We introduce a \textbf{pre-synthesis gate}: a protocol for \emph{counterfactual sensitivity faithfulness} that tests whether mutating high-saliency positions changes model output more than composition-matched controls. Cross-dataset transfer reveals two failure modes that would otherwise go undetected: \emph{faithful-but-wrong} (saliency valid, predictions fail) and \emph{inverted saliency} (top-saliency edits less impactful than random). Strikingly, models trained on mRNA-level assays collapse on a luciferase reporter dataset, demonstrating that protocol shifts can silently invalidate deployment. Across four benchmarks, 19/20 fold instances pass; the single failure shows inverted saliency. A biology-informed regularizer (BioPrior) strengthens saliency faithfulness with modest, dataset-dependent predictive trade-offs. Our results establish saliency validation as essential pre-deployment practice for explanation-guided therapeutic design. Code is available at \url{https://github.com/shadi97kh/BioPrior}.
\end{abstract}

\begin{figure}[t]
    \centering
    \includegraphics[width=\textwidth]{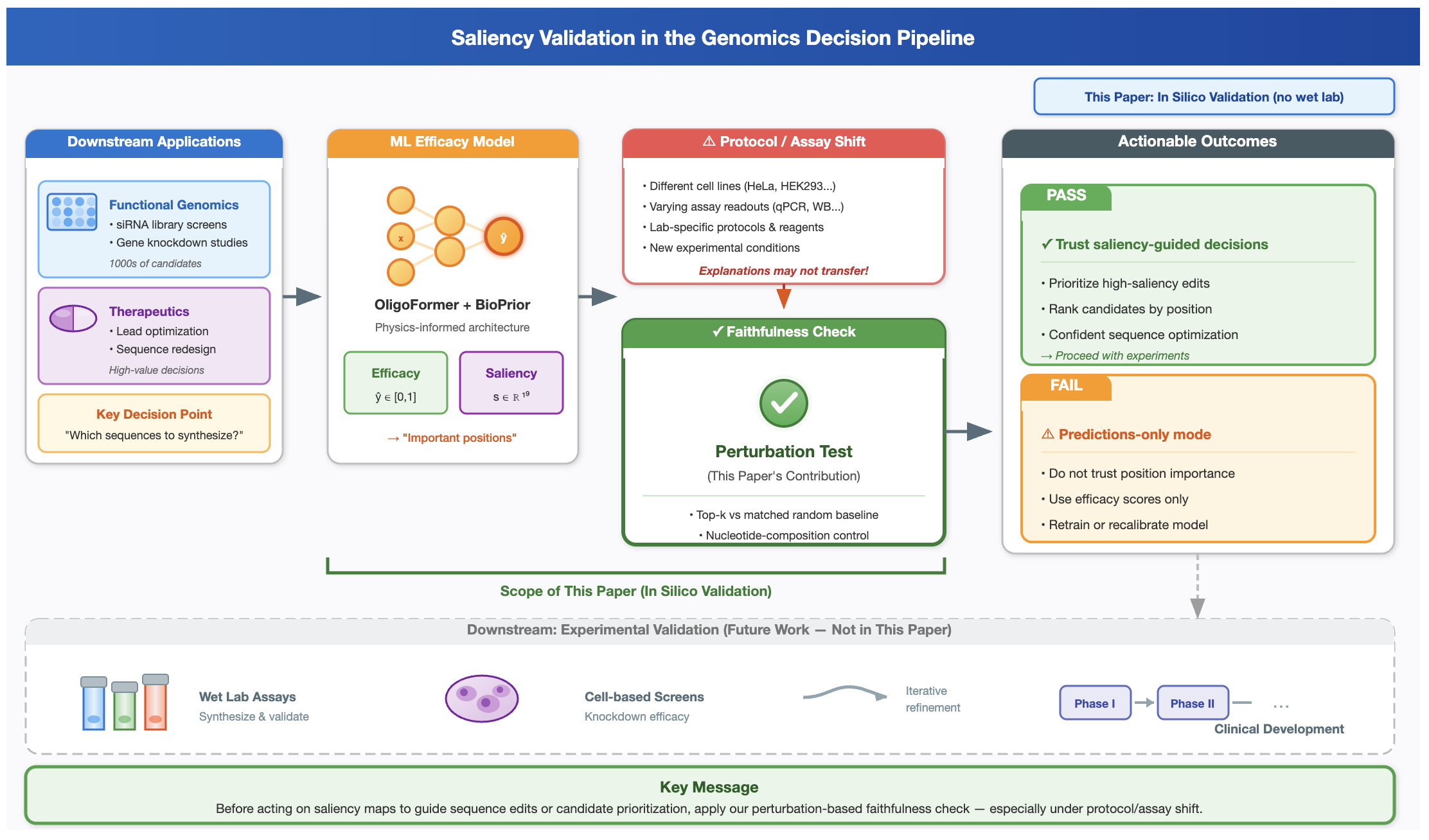}
    \caption{\textbf{Positioning saliency validation in the lab-in-the-loop decision pipeline.} In both therapeutic lead selection and functional genomics knockdown screens, researchers rely on predicted efficacy and position-level saliency (``important positions'') to decide which siRNA sequences to synthesize or prioritize for experimental validation. However, explanation methods can appear plausible while failing basic perturbation tests, a risk that compounds under assay or protocol shift across laboratories, cell lines, and readout technologies. This paper introduces a standardized faithfulness check (expected-effect perturbations with a nucleotide-matched baseline) that practitioners can apply as a \textbf{pre-synthesis gate} before acting on saliency maps in a new dataset or experimental setting. When validation passes, saliency-guided decisions (sequence edits, candidate ranking) can be trusted; when it fails, predictions may still be useful but position-importance reasoning should be avoided. Downstream wet-lab validation and clinical development (dashed region) are outside the scope of this work.}
    \label{fig:pipeline}
\end{figure}

\section{Introduction}
\label{sec:introduction}

Small interfering RNAs (siRNAs) enable programmable, sequence-specific gene silencing and have become a practical modality in both therapeutic development and functional genomics~\citep{elbashir2001duplexes, fire1998potent}. The clinical success of FDA-approved siRNA drugs, including patisiran, givosiran, and inclisiran~\citep{adams2018patisiran, balwani2020givosiran, ray2020inclisiran, setten2019development}, has intensified interest in computational methods for predicting siRNA efficacy. In discovery settings, researchers routinely screen many candidate oligonucleotides and prioritize those expected to achieve strong knockdown. This has driven sustained interest in machine learning models that predict siRNA efficacy directly from nucleotide sequence and related descriptors~\citep{han2018sirna, bai2024oligoformer}. In practice, this means: pick sequences, edit motifs or seed composition, adjust GC balance, and re-screen, so explanation quality directly affects experimental cost and iteration speed. Reliable saliency maps would enable practitioners to rationally edit candidate siRNA sequences at positions most likely to improve knockdown, accelerating the design of effective therapeutic oligonucleotides while reducing costly experimental iterations.

Modern deep predictors can be accurate on standard benchmarks, but an increasingly important question is whether they are \emph{trustworthy} as decision-support tools. In practice, investigators do not only use predicted efficacy scores; they often inspect saliency maps or other attribution visualizations to infer which nucleotides ``matter,'' and then use those attributions to motivate sequence edits (e.g., adjusting seed composition, GC balance, or motif avoidance). If these explanations are not faithful (meaning interventions at highlighted positions do not produce larger prediction changes than controls), then explanation-guided design can be misleading~\citep{adebayo2018sanity, kindermans2019reliability}, especially under the protocol and distribution shifts that are common across assays, labs, and readouts.

We therefore focus on a concrete, testable desideratum: \emph{a saliency method is useful for design only if mutating high-saliency positions changes the model's prediction more than mutating appropriate controls}~\citep{samek2017evaluating}. We term this \emph{counterfactual faithfulness}: the saliency map correctly identifies positions where the model is sensitive to interventions. We propose this test as a \textbf{pre-synthesis gate}, a validation step practitioners should run before acting on saliency maps in explanation-guided siRNA design. Crucially, this is a \textbf{model-centric guarantee}: saliency tracks where the model is sensitive, not necessarily what biology truly cares about. This is distinct from \emph{biological causality} (whether position changes affect true efficacy) and from \emph{distributional faithfulness} (whether saliency reflects learned correlations). Our test validates model sensitivity under single-base perturbations, which is the operationally relevant property for explanation-guided sequence editing.

We introduce a perturbation-based validation protocol that operationalizes this idea for nucleotide sequence predictors. Given a trained model and a held-out siRNA, we (i) compute position-wise saliency on nucleotide identity channels~\citep{simonyan2014deep, sundararajan2017axiomatic}, (ii) select the top-$k$ salient positions, (iii) quantify an expected-effect mutation score by averaging the prediction change under all single-base substitutions at those positions, and (iv) compare this score to a nucleotide-matched random baseline to control for compositional bias. This yields a simple pass/fail faithfulness test that can be run before saliency maps are used for design guidance~\citep{amorim2023evaluating}. Our protocol's key innovations over standard in-silico mutagenesis and faithfulness metrics are detailed in Section~\ref{sec:related} and Table~\ref{tab:ism_comparison}.

In parallel, we propose a biology-informed hybrid model for siRNA efficacy prediction that incorporates established design principles as differentiable regularizers. These priors include thermodynamic asymmetry~\citep{schwarz2003asymmetry}, seed-region composition constraints~\citep{jackson2010recognizing}, global GC heuristics~\citep{reynolds2004rational, uitei2004guidelines}, immune-motif avoidance~\citep{judge2005immunostimulatory}, and a duplex stability proxy penalizing excessive GC content that may indicate inaccessible targets. This approach aligns with recent work on physics-informed machine learning~\citep{raissi2019physics, karniadakis2021physics} and biologically interpretable neural networks~\citep{elmarakeby2021biologically, novakovsky2023obtaining, chen2024interpretable}, though biological systems present unique challenges compared to physical systems due to uncertain priors and context-dependent mechanisms~\citep{martinelli2025biology}.

Across four benchmark datasets, we find that saliency faithfulness holds in 95\% of fold–dataset settings and that salient positions cluster in canonical functional regions (seed and 3$'$ end). However, cross-dataset transfer reveals strong protocol dependence: models transfer among three datasets but fail on a luciferase-reporter dataset, suggesting assay-specific confounds that can break generalization even when within-dataset explanations are faithful. Our central thesis is that \textbf{saliency should be treated as a deploy-time claim}: validate explanation faithfulness on the target assay before using it to guide siRNA selection or modification.

\paragraph{Contributions.}
\begin{enumerate}
    \item \textbf{Introduce} a composition-controlled, perturbation-based protocol to validate saliency faithfulness for nucleotide sequence predictors, positioned as a pre-synthesis gate in lab-in-the-loop design workflows.
    
    \item \textbf{Demonstrate} that validated saliency aligns with biologically meaningful siRNA regions across multiple benchmarks (19/20 fold--dataset combinations pass), with high-saliency positions clustering at the 5$'$ and 3$'$ termini adjacent to known functional determinants.
    
    \item \textbf{Characterize} two distinct transfer failure modes: \emph{faithful-but-wrong} (saliency valid, predictions fail) and \emph{inverted saliency} (high-saliency positions less important than random), highlighting when explanations should not be trusted without dataset-specific validation.
    
    \item \textbf{Show} that a biology-informed regularizer (BioPrior) strengthens explanation faithfulness with dataset-dependent predictive effects, demonstrating that mechanism-informed training can improve saliency reliability.
\end{enumerate}

We release code and the validation protocol to enable adoption across sequence modeling applications.\footnote{Code available at \url{https://github.com/shadi97kh/BioPrior}.}

\section{Related Work}
\label{sec:related}

\paragraph{siRNA efficacy prediction: from rules to deep learning.}
Early siRNA design relied on empirically derived rules capturing thermodynamic asymmetry, seed-region composition, and GC-related heuristics \citep{uitei2004guidelines,reynolds2004rational,amarzguioui2004algorithm}.
More recent approaches learn nonlinear sequence-to-efficacy mappings using deep neural networks, including transformer-based models that integrate thermodynamic descriptors and pretrained RNA sequence representations for improved accuracy and transfer \citep{han2018sirna,bai2024oligoformer}, graph neural networks that capture siRNA-mRNA interaction dynamics \citep{long2024sirnadiscovery}, and preference-based ranking frameworks that address dataset bias through debiased training objectives \citep{zhang2025sidpt}.

\paragraph{Biology-informed regularization for sequence models.}
In scientific ML, incorporating domain knowledge as soft constraints or regularizers can improve robustness and align learned behavior with known mechanisms \citep{raissi2019physics,karniadakis2021physics}.
However, biological systems present unique challenges compared to physical systems: uncertain and context-dependent prior knowledge, heterogeneous and noisy data, partial observability, and complex high-dimensional networks \citep{martinelli2025biology}.
We adopt a biology-informed approach for siRNA prediction by encoding established design principles as differentiable penalties rather than hard constraints, allowing the model to learn dataset-specific deviations from canonical rules while remaining grounded in mechanistic understanding.
This aligns with recent work on biologically interpretable neural networks that maintain both predictive power and scientific transparency \citep{elmarakeby2021biologically,novakovsky2023obtaining,chen2024interpretable}.

\begin{figure*}[t]
    \centering
    \includegraphics[width=0.90\textwidth]{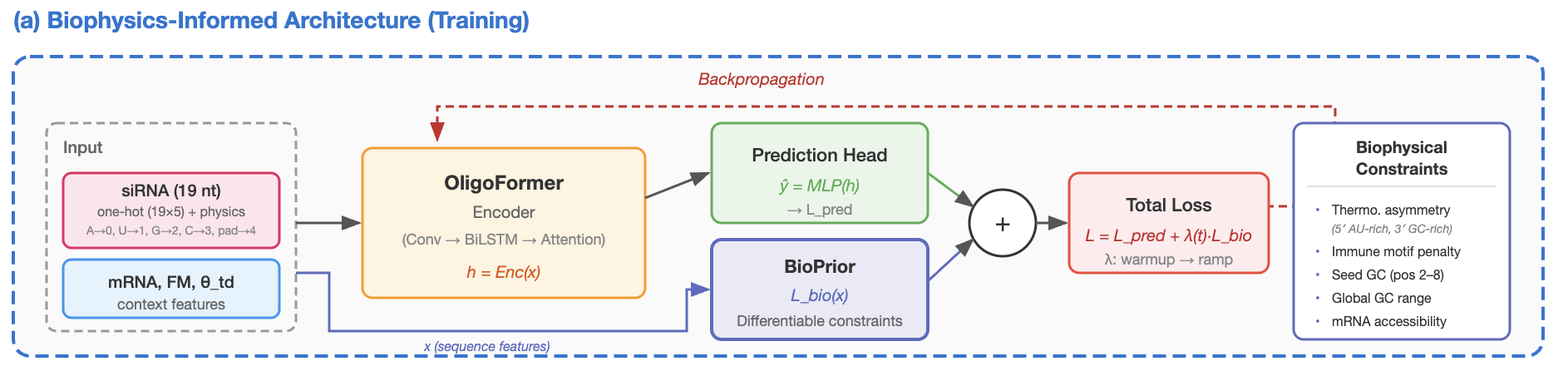}
    \vspace{0.3em}
    \includegraphics[width=0.90\textwidth]{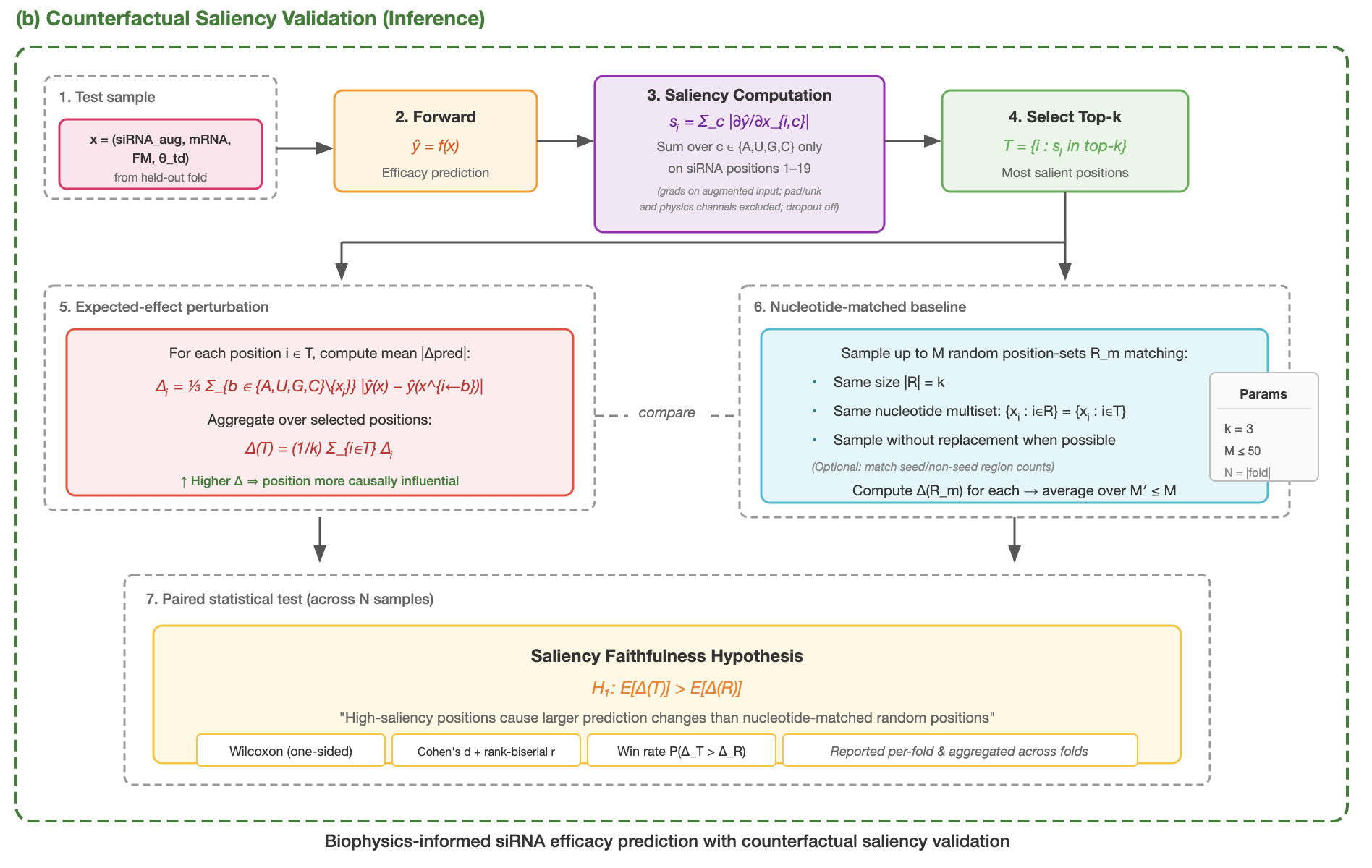}
    \caption{\textbf{Overview of training and saliency faithfulness.}
\textbf{(a)} A hybrid Conv--BiLSTM--Transformer with dual siRNA$\leftrightarrow$mRNA cross-attention predicts efficacy from sequence encodings plus RNA-FM and thermodynamic features, regularized by BioPrior constraints weighted by a schedule $\lambda(t)$. 
\textbf{(b)} Saliency is validated by perturbing top-$k$ salient siRNA positions (A/U/G/C channels only) and comparing the expected prediction change to a nucleotide-matched random baseline.}
    \label{fig:method_overview}
\end{figure*}

\paragraph{Saliency methods and faithfulness validation.}
Gradient-based saliency methods attribute predictions to input features via derivatives and variants such as integrated gradients \citep{simonyan2014deep,sundararajan2017axiomatic}.
However, saliency maps are not guaranteed to reflect true feature importance \citep{adebayo2018sanity,kindermans2019reliability,rudin2019stop}; a standard validation approach is perturbation-based testing, which measures whether modifying high-attribution features produces larger output changes than modifying suitable controls \citep{samek2017evaluating,amorim2023evaluating}.

\paragraph{Interpretability validation in sequence biology.}
In genomics, \emph{in-silico mutagenesis} (ISM) interprets sequence models by mutating positions and measuring prediction changes \citep{alipanahi2015predicting,zhou2018deep}.
Related approaches include motif discovery from learned filters \citep{lanchantin2017deep}, TF-MoDISco \citep{shrikumar2018technical}, and deletion/insertion metrics \citep{hooker2019benchmark}.
Our protocol differs from both ISM and standard faithfulness evaluation \citep{samek2017evaluating,hooker2019benchmark} in four ways: (1) an expected-effect operator averaging over all 3 substitutions per position; (2) a composition-matched baseline controlling for nucleotide-specific sensitivity; (3) explicit pass/fail acceptance criteria for deploy-time decisions; and (4) a cross-dataset diagnostic taxonomy (faithful-but-wrong vs.\ inverted-saliency). These components are individually known but their combination for nucleotide sequences is novel; Table~\ref{tab:ism_comparison} summarizes the ISM comparison.

\section{Background}
\label{sec:background}

\paragraph{Task and datasets.}
We study siRNA efficacy prediction from 19-mer guide sequences, where efficacy $y \in [0,1]$ reflects target knockdown strength. We index siRNA positions 1--19 from 5$'$ to 3$'$ throughout this paper.
We evaluate on four public benchmarks spanning distinct experimental protocols and distribution shift (Hu, Taka, Mix, Shabalina; Table~\ref{tab:datasets_full}), and normalize efficacy scores to $[0,1]$.
Following common practice, we define the high-efficacy regime as $y \ge 0.7$ and report both continuous (correlation) and thresholded (AUC) metrics.

\paragraph{Gradient-based saliency.}
Given a trained predictor $f_\theta(\mathbf{x})$ and an siRNA input representation with nucleotide channels $\mathbf{X}^{\text{si}} \in \mathbb{R}^{19 \times C}$, we compute position-wise saliency using the gradient magnitude restricted to nucleotide identity (A/U/G/C) channels:
\begin{equation}
s_i \;=\; \sum_{c \in \{A,U,G,C\}} \left| \frac{\partial f_\theta(\mathbf{x})}{\partial \mathbf{X}^{\text{si}}_{i,c}} \right|.
\label{eq:saliency_bg}
\end{equation}

\paragraph{Faithfulness desideratum.}
A saliency map is \emph{faithful} if perturbing high-saliency positions causes larger prediction changes than perturbing appropriate controls \citep{samek2017evaluating}.
We operationalize this via counterfactual single-nucleotide substitutions and compare the expected prediction change for top-$k$ salient positions against a nucleotide-matched random baseline (Section~\ref{sec:saliency_validation}).

\paragraph{Faithfulness taxonomy.} To avoid confusion, we distinguish three notions:

\begin{center}
\small
\begin{tabular}{lll}
\toprule
\textbf{Term} & \textbf{Definition} & \textbf{Validated by} \\
\midrule
Sensitivity faithfulness & High-saliency positions change model output & Perturbation test (this work) \\
& more than matched controls & \\
Causal faithfulness & Interventions change ground-truth efficacy & Wet-lab experiments \\
Stability & Explanations invariant to irrelevant transforms & Input perturbation \\
\bottomrule
\end{tabular}
\end{center}

\noindent Our test validates \textbf{sensitivity faithfulness}, the operationally relevant property for explanation-guided sequence editing. We also introduce terminology for transfer failures:
\begin{itemize}[leftmargin=*,itemsep=1pt,topsep=2pt]
    \item \textbf{Faithful-but-wrong:} Saliency test passes but predictions fail (model is internally consistent but learned wrong rules).
    \item \textbf{Inverted saliency:} Saliency test fails with $d_z < 0$ (high-saliency positions are \emph{less} important than random).
\end{itemize}

\begin{table}[h]
\centering
\small
\begin{tabular}{lcc}
\toprule
& \textbf{ISM} & \textbf{Ours} \\
\midrule
Purpose & Interpretation & Validation \\
Output & Mutation effect profile & Pass/fail decision \\
Baseline & None (raw effects) & Composition-matched \\
Statistical test & Optional & Required (Wilcoxon) \\
Intended use & Post-hoc explanation & Deploy-time gate \\
\bottomrule
\end{tabular}
\caption{Comparison: In-silico mutagenesis (ISM) vs.\ our protocol.}
\label{tab:ism_comparison}
\end{table}

\noindent Unlike ISM, which is an interpretability \emph{output}, our protocol is a statistical \emph{acceptance test} that decides whether explanations are safe to use for editing under a given dataset.

\section{Method}
\label{sec:method}

We propose a biology-informed architecture for siRNA efficacy prediction that integrates established design principles as differentiable regularizers and introduces a counterfactual perturbation protocol to validate gradient-based saliency. Concretely, our method extends OligoFormer-style sequence modeling with (i) a hybrid encoder combining convolution, BiLSTM, and self-attention, (ii) dual-stream siRNA--mRNA cross-attention fusion, (iii) a BioPrior module that computes differentiable constraint penalties from sequence-derived features (optionally conditioned on per-position nucleotide probabilities), and (iv) a saliency faithfulness test based on an expected-effect perturbation operator with a nucleotide-matched random baseline.

\paragraph{Overview.}
Figure~\ref{fig:method_overview} summarizes our approach.
During training (Fig.~\ref{fig:method_overview}a), we optimize a predictive model while softly enforcing established siRNA design principles via a differentiable BioPrior regularizer whose contribution is scheduled over epochs.
During evaluation (Fig.~\ref{fig:method_overview}b), we validate whether gradient-based saliency exhibits \emph{counterfactual faithfulness} using a perturbation test that compares the effect of mutating top-saliency positions against a nucleotide-matched baseline.

\subsection{Model Architecture}
\label{sec:architecture}

\paragraph{Overview.}
Our backbone follows \textbf{OligoFormer} \citep{bai2024oligoformer}: a hybrid \textbf{Conv $\rightarrow$ BiLSTM $\rightarrow$ attention} encoder for siRNA and mRNA, followed by \textbf{bidirectional cross-attention} and an MLP prediction head. The appendix provides full architectural hyperparameters and implementation details (Appendix~\ref{sec:implementation}).

\paragraph{Inputs.}
Each example contains (i) a 19-nt siRNA guide, (ii) an mRNA context window centered at the target site, and (iii) global descriptors.
We encode siRNA as $\mathbf{X}^{\text{si}} \in \mathbb{R}^{19 \times C_{\text{si}}}$ where $C_{\text{si}}=13$ includes a 5-channel one-hot (A,U,G,C,pad) plus lightweight sequence-derived indicator channels (e.g., seed/cleavage flags and AU/GC indicators).
We encode mRNA as $\mathbf{X}^{\text{mr}} \in \mathbb{R}^{L_{\text{mr}} \times 5}$ (A,U,G,C,pad).
We additionally use pooled RNA-FM embeddings $\mathbf{z}_{\text{FM}}^{\text{si}}, \mathbf{z}_{\text{FM}}^{\text{mr}}$ \citep{chen2022rnafm} and a thermodynamic descriptor vector $\mathbf{z}_{\text{td}} \in \mathbb{R}^{d_{\text{td}}}$ computed from sequence and thermodynamic tools~\citep{lorenz2011viennarna}.

\paragraph{Encoders and fusion.}
Both siRNA and mRNA are encoded with a shallow convolutional front-end, a 2-layer BiLSTM~\citep{hochreiter1997lstm}, and a lightweight self-attention block~\citep{vaswani2017attention}, producing contextual representations $\tilde{\mathbf{H}}^{\text{si}}$ and $\tilde{\mathbf{H}}^{\text{mr}}$.
We then fuse the two streams using \textbf{dual cross-attention} (siRNA$\leftarrow$mRNA and mRNA$\leftarrow$siRNA), pool each stream by mean+max pooling, concatenate global descriptors $(\mathbf{z}_{\text{FM}}^{\text{si}}, \mathbf{z}_{\text{FM}}^{\text{mr}}, \mathbf{z}_{\text{td}})$, and map the result to the final efficacy prediction via an MLP.

\paragraph{Per-position nucleotide probability head.}
To couple the predictor to differentiable mechanistic constraints, we attach a small auxiliary head that outputs per-position nucleotide probabilities
$\mathbf{P}^{\text{si}} \in [0,1]^{19\times 4}$ from intermediate siRNA features (after the BiLSTM encoder).
These probabilities are computed via softmax over learned logits $\mathbf{z} \in \mathbb{R}^{19 \times 4}$:
\[
P_{i,b} = \frac{\exp(z_{i,b})}{\sum_{b' \in \{A,U,G,C\}} \exp(z_{i,b'})}
\]
The probabilities are used only by BioPrior (below) and do not alter the main prediction pathway.

\subsection{Biology-Informed Regularization (BioPrior)}
\label{sec:constraints}

We encode established siRNA design principles as differentiable penalties computed from deterministic sequence-derived features and optional soft nucleotide probabilities $\mathbf{P}^{\text{si}}$.
The BioPrior module aggregates constraint losses:
\begin{equation}
\mathcal{L}_{\text{bio}} = \sum_{c \in \mathcal{C}} \bar{\alpha}_c \, \mathcal{L}_{c},
\end{equation}
where $\mathcal{C}$ includes (i) terminal asymmetry preference, (ii) seed composition constraints, (iii) global GC constraints, (iv) immune motif avoidance, and (v) a duplex stability proxy based on siRNA GC content.
We compute BioPrior on explicit sequence-level quantities (rather than latent representations) to avoid injecting spurious position biases.
A full mathematical specification and proofs are provided in Appendix~\ref{app:bioprior}.

\paragraph{Scheduling.}
We introduce BioPrior with an epoch-based warmup-and-ramp:
\begin{equation}
\lambda(t) =
\begin{cases}
0 & t < t_{\text{warm}},\\
\min(\lambda_{\max}, \lambda_0 + \gamma (t - t_{\text{warm}})) & \text{otherwise},
\end{cases}
\end{equation}
with $t_{\text{warm}} = 8$ epochs, $\lambda_0 = 0.10$, $\gamma = 0.01$ per epoch, and $\lambda_{\max} = 0.30$.
This schedule begins at $\lambda_0 = 0.10$ after warmup and ramps linearly, allowing the model
to first learn predictive features before gradually increasing biological regularization.
The relatively high cap accommodates the normalized constraint weights
(Section~\ref{sec:constraints}), which distribute the effective penalty across five
terms; the per-constraint contribution remains moderate throughout training.

\subsection{Training Objective}
\label{sec:training}

The overall objective combines prediction, BioPrior regularization, and auxiliary mechanistic supervision:
\begin{equation}
\mathcal{L}_{\text{total}} = \mathcal{L}_{\text{pred}} + \lambda(t)\,\mathcal{L}_{\text{bio}} + \lambda_{\text{aux}}\,\mathcal{L}_{\text{aux}}.
\end{equation}

\textbf{Gradient flow through soft probabilities.} The model produces soft nucleotide probability distributions $P \in \mathbb{R}^{L \times 4}$ (where $P_{i,b} = \Pr(\text{position } i = b)$, $\sum_b P_{i,b} = 1$) alongside efficacy predictions. \textbf{In all main experiments}, BioPrior penalties are computed on these soft probabilities: $\mathcal{L}_{\text{bio}} = \mathcal{L}_{\text{bio}}(P_{\text{model}})$. This ensures gradients from biological constraints backpropagate through the model's sequence representation. At inference with fixed sequences, one-hot encodings are used (gradient flow not needed). An ablation using deterministic one-hot BioPrior is in Appendix~\ref{app:full_results}.

We optimize with AdamW and standard regularization/early stopping under 5-fold cross-validation; the complete optimization protocol, hyperparameters, and compute details are listed in Appendix~\ref{sec:implementation}.

\subsection{Counterfactual Saliency Validation}
\label{sec:saliency_validation}
\label{sec:perturbation} 

We test whether gradient saliency highlights positions with high interventional sensitivity using a counterfactual perturbation protocol with a nucleotide-matched baseline.
Unless stated otherwise, we compute saliency using the gradient magnitude restricted to the siRNA nucleotide identity channels (A/U/G/C), aligning attribution with the intervention space used in our perturbation test; Integrated Gradients is reported only as a robustness check (Appendix~\ref{app:sanity}).
Specifically, saliency is computed with respect to the one-hot nucleotide channels at the model input layer; gradients flow through any learned projections in the standard manner. Derived channels (GC content, seed indicators, thermodynamic features) are excluded from attribution because they are deterministically recomputed after perturbation, so attributing to nucleotide identity captures the causal parent in the input graph.

\textbf{Important:} Gradient-based saliency measures model sensitivity to nucleotide channels \emph{holding derived features fixed}; it is an approximation. The perturbation validation (Section~\ref{sec:perturbation}) is our primary interpretability test because it recomputes all derived features after each mutation, capturing the true causal effect of nucleotide changes.

Given a held-out example, we compute position saliency $s_i$ via Eq.~\ref{eq:saliency_bg}, select the top-$k$ positions $T$, and quantify their \emph{expected effect} under all single-base substitutions:
\begin{equation}
\Delta_i = \frac{1}{3} \sum_{b \in \{A,U,G,C\}\setminus\{x_i\}}
\left| \hat{y}(\mathbf{X}) - \hat{y}(\mathbf{X}^{i \leftarrow b}) \right|, 
\qquad
\Delta(T) = \frac{1}{k}\sum_{i\in T}\Delta_i.
\end{equation}
After each substitution we recompute all derived input channels (seed indicators, GC content, thermodynamic asymmetry) to ensure input coherence; this deterministic recomputation can introduce nonlinear effects that break simple gradient-effect correspondence.

\textit{What is recomputed per mutation:} Derived indicator channels and thermodynamic descriptors are recomputed after each substitution. RNA-FM embeddings are held fixed for tractability; we verify on a 50-sequence subset that recomputing FM changes $d_z$ by $<0.05$ (Appendix~\ref{app:sanity}).
We test faithfulness under single-base edits, the operative action in sequence refinement, rather than distribution-preserving perturbations.

\paragraph{Averaging structure.} Our test averages over three levels: (i) 3 substitutions per position (all bases except the original), (ii) $k$ top-saliency positions, and (iii) $M'$ composition-matched random sets. This yields one paired difference $d = \Delta(T) - \Delta_{\text{match}}$ per sequence, enabling standard paired statistical testing across the held-out set.

\paragraph{Why this test is nontrivial.} Gradient-based saliency and perturbation effects are not guaranteed to align:
\begin{itemize}[leftmargin=*,itemsep=1pt,topsep=2pt]
    \item \textbf{Discrete perturbations:} We perturb nucleotide identities, not continuous features; local linearity assumptions may fail.
    \item \textbf{Derived channel recomputation:} After each mutation we recompute seed indicators, GC content, and thermodynamic features, producing nonlinear, discontinuous changes that gradients do not capture.
    \item \textbf{Negative controls fail:} Randomized-weight, shuffled-label, shuffled-saliency, and bottom-$k$ controls all fail the test (Table~\ref{tab:negative_controls}).
    \item \textbf{Transfer produces inverted saliency:} Taka$\to$Hu yields $d_z = -1.25$, where high-saliency positions are \emph{less} sensitive than random. This smoking-gun failure confirms the test has discriminative power.
\end{itemize}
Additional details on gradient saturation and intervention-space alignment are in Appendix~\ref{app:sanity}.

To control composition bias, we sample random position sets with the \emph{same nucleotide multiset} as $T$ and compute the matched baseline $\Delta(R)$.
We additionally run a stricter \emph{region+composition matched} baseline; faithfulness remains significant under this control (Table~\ref{tab:region_matched} in Appendix~\ref{app:region_matched}).
We evaluate faithfulness across held-out samples using paired one-sided Wilcoxon signed-rank tests~\citep{wilcoxon1945individual} and report effect sizes.

\paragraph{Limitation: position vs.\ composition matching.} Our primary baseline matches nucleotide composition but not positional region. If high-saliency positions cluster at inherently sensitive regions (e.g., sequence ends), this could inflate results. To address this concern, we conducted additional experiments with a \emph{region-matched} baseline that samples random positions from the same coarse region buckets as top-$k$ (5$'$ terminus: 1--4, seed-adjacent: 5--8, cleavage: 9--11, mid: 12--15, 3$'$ terminus: 16--19) while also matching nucleotide composition. Results remain significant: Hu achieves 78.3\% win rate ($d_z = 0.71$) under region+composition matching versus 85.2\% ($d_z = 0.86$) under composition-only matching. The reduced but still significant effect confirms that saliency captures position-specific patterns beyond regional sensitivity.

Additionally, our transfer analysis provides indirect evidence against positional bias: Taka-trained models show high saliency at positions 9--11, yet these positions show \emph{inverted} sensitivity when tested on other datasets. If positional bias were dominant, we would expect consistent sensitivity regardless of training source. The asymmetric transfer failures suggest saliency captures learned position-specific patterns rather than inherent regional sensitivity.

The full mathematical specification is provided in Appendix~\ref{app:perturbation}.

\begin{table}[t]
\centering
\small
\fbox{
\parbox{0.95\columnwidth}{
\textbf{Algorithm: Counterfactual Faithfulness Test}
\vspace{0.3em}

\textbf{Input:} Model $f_\theta$, held-out set $\mathcal{D}$, top-$k$, matched samples $M'$

\textbf{Output:} Pass/fail, effect size $d_z$, win rate

\vspace{0.2em}
\texttt{for} each $(\mathbf{x}, y) \in \mathcal{D}$:
\begin{enumerate}[leftmargin=1.5em,itemsep=0pt,topsep=2pt]
    \item Compute saliency: $s_i = \sum_{c} |\frac{\partial \hat{y}}{\partial x_{i,c}}|$
    \item Select top-$k$ positions: $T = \text{argtop}_k(s)$
    \item Compute expected effect: $\Delta(T) = \frac{1}{k}\sum_{i \in T} \frac{1}{3}\sum_{b \neq x_i} |\hat{y}(\mathbf{x}) - \hat{y}(\mathbf{x}^{i \leftarrow b})|$
    \item Sample $M'$ composition-matched random sets $\{R_m\}$
    \item Compute baseline: $\Delta_{\text{match}} = \frac{1}{M'}\sum_m \Delta(R_m)$
    \item Record difference: $d = \Delta(T) - \Delta_{\text{match}}$
\end{enumerate}
\texttt{end for}

\vspace{0.2em}
Compute Wilcoxon signed-rank $p$, Cohen's $d_z$, win rate

\textbf{Pass} if: $p < 0.05$ \textbf{and} $d_z > 0.2$ \textbf{and} win rate $> 50\%$
}}
\end{table}

\begin{table}[t]
\centering
\small
\fbox{
\parbox{0.95\columnwidth}{
\textbf{Faithfulness Test Specification}
\vspace{0.3em}

\textit{Test:} Paired one-sided Wilcoxon signed-rank; $H_1$: $\mathbb{E}[\Delta(T)] > \mathbb{E}[\Delta_{\text{match}}]$

\textit{Defaults:} $k=3$, $N \geq 100$ samples, $M'=50$ matched sets per sample

\textit{Pass criteria:} $p < 0.05$ \textbf{and} $d_z > 0.2$ \citep{cohen1988statistical} \textbf{and} win rate $> 50\%$

\textit{Threshold calibration:} Trained models achieve $d_z = 0.70$--$1.07$; negative controls yield $d_z \in [-0.45, 0.03]$. The $d_z > 0.2$ threshold cleanly separates learned from spurious saliency.

\textit{Scope:} Validates model sensitivity (where edits change predictions), \textbf{not} biological causality or OOD generalization.

\textit{If fail:} Do not use saliency for design; check for distribution shift; retrain on protocol-matched data.
}}
\end{table}

\section{Experiments}
\label{sec:experiments}

We evaluate our biology-informed sequence model on four benchmark siRNA datasets, assessing (i) predictive performance and (ii) explanation faithfulness using our counterfactual saliency validation protocol.

\subsection{Datasets}
\label{sec:datasets}

We use four publicly available siRNA efficacy datasets spanning different experimental protocols and assays (Table~\ref{tab:datasets_full}).

\paragraph{Huesken (Hu).} The largest benchmark dataset, containing 2,431 siRNAs targeting 34 human genes measured in H1299 cells \citep{huesken2005design}. Efficacy was quantified via branched DNA assay measuring residual mRNA levels 24 hours post-transfection. Scores are normalized to $[0,1]$ where higher values indicate stronger knockdown.

\paragraph{Katoh (Taka).} A dataset of 702 siRNAs all targeting a single luciferase reporter mRNA, evaluated using dual-luciferase reporter assays in HeLa cells \citep{katoh2007sirna}. This dataset exhibits different distributional properties and position-importance patterns compared to Hu, likely reflecting differences in assay design and the single-target nature of the experiments.

\paragraph{Mix.} Following \citet{bai2024oligoformer}, we adopt the dataset categorization from OligoFormer, which collected nine datasets comprising 3,714 siRNAs and 75 mRNAs from prior studies, then organized them into three groups: the Huesken dataset, the Takayuki dataset, and the remaining seven studies merged into a single ``Mixset.'' The resulting Mix dataset contains 581 siRNAs compiled from these seven independent studies with heterogeneous experimental protocols: Amarzguioui (46 siRNAs targeting 4 mRNAs in HaCaT cells) \citep{amarzguioui2003tolerance}, Harborth (44 siRNAs targeting 1 mRNA in HeLa) \citep{harboth2003sequence}, Hsieh (108 siRNAs targeting 22 mRNAs in HEK293T) \citep{hsieh2004library}, Khvorova (14 siRNAs targeting 1 mRNA in HEK293) \citep{khvorova2003functional}, Reynolds (240 siRNAs targeting 7 mRNAs in HEK293) \citep{reynolds2004rational}, Vickers (76 siRNAs targeting 2 mRNAs in T24) \citep{vickers2003efficient}, and Ui-Tei (53 siRNAs targeting 3 mRNAs in HeLa) \citep{uitei2004guidelines}. Inhibition efficacy across all constituent datasets was normalized to $[0,1]$, with 0.7 used as the threshold for high-efficacy classification, consistent with the OligoFormer preprocessing. Needleman--Wunsch global alignment \citep{needleman1970general} was applied to remove redundant sequences (identity $>80\%$) across splits, yielding the final dataset. While the aggregation provides scale and diversity across cell lines and target genes, the mixed provenance introduces additional noise and potential batch effects.

\paragraph{Shabalina.} A curated dataset of 653 siRNAs aggregated from published literature with computational filtering for thermodynamic properties \citep{shabalina2006computational}. The curation process may introduce selection biases favoring well-characterized sequences.

\paragraph{Threshold sensitivity.} Because datasets are independently normalized to $[0,1]$, the 0.7 threshold may correspond to different absolute knockdown levels across assays. We verified that transfer conclusions are robust to threshold choice: AUC rankings remain consistent at 0.6, 0.7, and 0.8 thresholds, and using top-30\% per dataset as positives yields the same transfer patterns (Taka remains the outlier). Full sensitivity analysis is in Appendix~\ref{app:full_results}.

\begin{table}[t]
\centering
\small
\begin{tabular}{lccccccc}
\toprule
\textbf{Dataset} & \textbf{$N$} & \textbf{Targets} & \textbf{Sources} & \textbf{Mean} & \textbf{Std} & \textbf{High-eff.} & \textbf{Assay} \\
\midrule
Hu & 2,431 & 34 & 1 & 0.58 & 0.23 & 847 (34.8\%) & bDNA \\
Taka & 702 & 1 & 1 & 0.61 & 0.25 & 298 (42.5\%) & Luciferase \\
Mix & 581 & 40 & 7 & 0.56 & 0.24 & 183 (31.5\%) & Mixed \\
Shabalina & 653 & $>$30 & 1 & 0.54 & 0.26 & 199 (30.5\%) & Literature \\
\bottomrule
\end{tabular}
\caption{Dataset characteristics. High-efficacy threshold is $y \geq 0.7$.}
\label{tab:datasets_full}
\end{table}

\subsection{Baselines and Ablations}
\label{sec:baselines}

\paragraph{OligoFormer baseline.} We compare to the published OligoFormer model \citep{bai2024oligoformer}, which combines sequence representations with thermodynamic and foundation-model features.

\paragraph{Rule-based baseline.} We include a classical thermodynamic scoring baseline based on established heuristic rules (e.g., Reynolds \citep{reynolds2004rational}, Ui-Tei \citep{uitei2004guidelines}).

\paragraph{Ablations.} To isolate contributions of our components, we evaluate:
\begin{itemize}[leftmargin=*,itemsep=2pt]
    \item \textbf{Baseline (no BioPrior):} identical architecture trained with prediction loss only (biology regularization disabled; $\lambda(t)=0$).
    \item \textbf{+BioPrior (scheduled):} full model with biology loss $\mathcal{L}_{\text{bio}}$ enabled and scheduled via warmup--ramp $\lambda(t)$.
\end{itemize}
When reporting saliency faithfulness, we use the same faithfulness protocol for all variants to ensure comparability.

\subsection{Evaluation Metrics}
\label{sec:eval_metrics}

\paragraph{Predictive performance.} We report:
\begin{itemize}[leftmargin=*,itemsep=2pt]
    \item \textbf{MSE}: mean squared error on $y\in[0,1]$.
    \item \textbf{Pearson $r$} and \textbf{Spearman $\rho$}: correlation between predicted and true efficacy.
    \item \textbf{ROC-AUC} and \textbf{PR-AUC}: treating $y\ge 0.7$ as the positive class.
\end{itemize}

\paragraph{Faithfulness.} Using the protocol in Section~\ref{sec:saliency_validation}, we report:
\begin{itemize}[leftmargin=*,itemsep=2pt]
    \item \textbf{Win rate}: fraction of samples where $\Delta(T)>\Delta_{\text{match}}$ (top-$k$ exceeds the nucleotide-matched baseline).
    \item \textbf{Paired effect sizes}: Cohen's $d_z$ (paired standardized mean difference: $d_z = \bar{d}/s_d$) and rank-biserial correlation on paired differences.
    \item \textbf{Statistical significance}: one-sided paired Wilcoxon signed-rank test for $H_1:\mathbb{E}[\Delta(T)]>\mathbb{E}[\Delta_{\text{match}}]$.
\end{itemize}

All metrics are reported as mean $\pm$ standard deviation across folds.

\section{Results}
\label{sec:results}

We present results on predictive performance, cross-dataset generalization, saliency faithfulness validation, and ablation studies. All experiments use 5-fold cross-validation; we report mean $\pm$ std across folds. Transfer experiments were repeated with 3 random seeds per configuration; we report aggregated statistics with complete tables in Appendix~\ref{app:transfer_faith}.

\subsection{Intra-Dataset Predictive Performance}
\label{sec:pred_results}

Table~\ref{tab:predictive} summarizes 5-fold cross-validation performance. Our biology-informed model (+BioPrior) achieves consistent improvements over the baseline: +0.01 AUC, +0.02 PR-AUC, and +0.01 PCC on average across datasets. At the fold level, BioPrior improves AUC in 15/20 fold-dataset combinations (75\%), with the largest gains on Taka (+0.02 AUC, +0.04 PR-AUC). BioPrior also strengthens saliency faithfulness (Section~\ref{sec:ablations}).

\begin{table}[t]
\centering
\small
\begin{tabular}{llcccc}
\toprule
\textbf{Dataset} & \textbf{Model} & \textbf{AUC} & \textbf{PR-AUC} & \textbf{PCC} & \textbf{F1} \\
\midrule
\multirow{2}{*}{Hu} 
& Baseline & 0.82{\tiny $\pm$.01} & 0.81{\tiny $\pm$.02} & 0.63{\tiny $\pm$.02} & 0.77{\tiny $\pm$.01} \\
& +BioPrior & \textbf{0.83}{\tiny $\pm$.02} & \textbf{0.82}{\tiny $\pm$.02} & \textbf{0.64}{\tiny $\pm$.02} & 0.77{\tiny $\pm$.03} \\
\midrule
\multirow{2}{*}{Mix} 
& Baseline & 0.80{\tiny $\pm$.04} & 0.80{\tiny $\pm$.06} & 0.60{\tiny $\pm$.04} & 0.76{\tiny $\pm$.06} \\
& +BioPrior & \textbf{0.81}{\tiny $\pm$.05} & \textbf{0.81}{\tiny $\pm$.07} & \textbf{0.61}{\tiny $\pm$.08} & \textbf{0.77}{\tiny $\pm$.06} \\
\midrule
\multirow{2}{*}{Taka} 
& Baseline & 0.82{\tiny $\pm$.05} & 0.63{\tiny $\pm$.10} & 0.68{\tiny $\pm$.08} & 0.59{\tiny $\pm$.07} \\
& +BioPrior & \textbf{0.84}{\tiny $\pm$.05} & \textbf{0.67}{\tiny $\pm$.10} & 0.68{\tiny $\pm$.08} & \textbf{0.60}{\tiny $\pm$.07} \\
\midrule
\multirow{2}{*}{Shabalina} 
& Baseline & 0.72{\tiny $\pm$.04} & 0.66{\tiny $\pm$.07} & 0.49{\tiny $\pm$.05} & 0.65{\tiny $\pm$.06} \\
& +BioPrior & 0.72{\tiny $\pm$.02} & 0.66{\tiny $\pm$.06} & 0.49{\tiny $\pm$.03} & \textbf{0.68}{\tiny $\pm$.03} \\
\bottomrule
\end{tabular}
\caption{Intra-dataset predictive performance (5-fold CV). Best results in \textbf{bold}. PR-AUC included given class imbalance (30--42\% positives).}
\label{tab:predictive}
\end{table}

\subsection{Cross-Dataset Generalization}
\label{sec:transfer_results}

A critical test of biological validity is whether models generalize across datasets collected under different experimental conditions. Table~\ref{tab:transfer} presents inter-dataset transfer results, revealing substantial heterogeneity in generalization patterns.

\begin{table}[t]
\centering
\small
\begin{tabular}{llccc}
\toprule
\textbf{Source} & \textbf{Target} & \textbf{Baseline} & \textbf{+BioPrior} & \textbf{Diff} \\
\midrule
\multicolumn{5}{l}{\textit{Strong generalization (AUC $> 0.75$)}} \\
Shabalina & Mix & 0.813 & \textbf{0.816} & +0.3\% \\
Mix & Hu & 0.792 & 0.792 & 0.0\% \\
Shabalina & Hu & 0.786 & \textbf{0.787} & +0.1\% \\
Hu & Mix & 0.775 & 0.773 & -0.2\% \\
\midrule
\multicolumn{5}{l}{\textit{Moderate generalization (AUC 0.65--0.75)}} \\
Mix & Shabalina & 0.713 & 0.712 & -0.1\% \\
Hu & Shabalina & 0.699 & 0.698 & -0.1\% \\
\midrule
\multicolumn{5}{l}{\textit{Poor generalization (AUC $< 0.60$)}} \\
Shabalina & Taka & 0.574 & 0.559 & -1.5\% \\
Hu & Taka & 0.536 & 0.535 & -0.1\% \\
Mix & Taka & 0.499 & 0.497 & -0.2\% \\
\midrule
\multicolumn{5}{l}{\textit{Failed transfer from Taka (inverted predictions)}} \\
Taka & Shabalina & 0.517 & 0.517 & 0.0\% \\
Taka & Mix & 0.510 & 0.510 & 0.0\% \\
Taka & Hu & 0.490 & 0.490 & 0.0\% \\
\bottomrule
\end{tabular}
\caption{Inter-dataset transfer performance (AUC). Models trained on source dataset, evaluated on target.}
\label{tab:transfer}
\end{table}

\paragraph{Key findings.} Three patterns emerge from the transfer experiments:

\textbf{Cross-dataset sequence overlap.} Before interpreting transfer results, we verified that sequence overlap does not inflate performance. Using Needleman-Wunsch alignment with 80\% identity threshold, we found minimal overlap: Hu-Mix share 12 sequences (0.5\%), Hu-Shabalina share 8 (0.3\%), and Taka shares $<$5 sequences with any other dataset. Transfer performance thus reflects genuine generalization rather than memorization of shared sequences.

\textbf{(1) Asymmetric generalization.} Transfer success is highly asymmetric. Shabalina$\to$Mix achieves 0.816 AUC while Mix$\to$Shabalina reaches only 0.713. This suggests dataset-specific biases that benefit transfer in one direction but not the reverse.

\begin{figure*}[t]
\centering
\includegraphics[width=\textwidth]{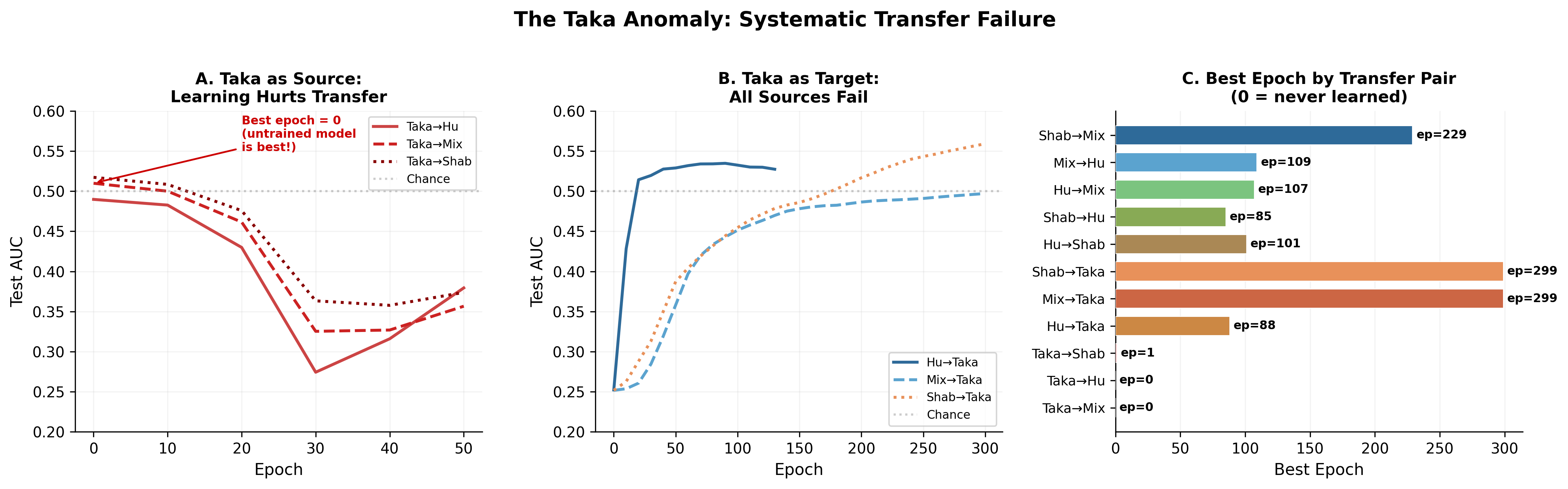}
\caption{\textbf{The Taka transfer anomaly.}
\textbf{(a)} When Taka is used as the training source, test AUC on all other datasets \emph{decreases} during training; the untrained model (epoch~0) transfers better than the trained model (best epoch = 0 for all three targets).
\textbf{(b)} When Taka is the target, all source models fail to exceed chance-level AUC regardless of training duration.
\textbf{(c)} Best epoch across all 11 mechanistic transfer pairs: successful transfers (Shab$\to$Mix, Mix$\to$Hu, etc.) converge at intermediate epochs, while all Taka-involving pairs either never learn (epoch 0) or train to exhaustion (epoch 299) without meaningful improvement. This asymmetry confirms that the Taka incompatibility is systematic and bidirectional.}
\label{fig:taka_anomaly}
\end{figure*}

\textbf{(2) Taka as systematic outlier.} Models trained on Taka fail to generalize to any other dataset (AUC $\approx 0.50$), and models trained on other datasets fail on Taka. The negative PCC values when Taka-trained models are applied elsewhere suggest inverted label relationships: sequences that Taka labels as high-efficacy may violate rules that predict efficacy in other datasets. Strikingly, Taka-trained models achieve their best transfer performance at epoch~0 (the untrained model), with test AUC \emph{decreasing} during training (Figure~\ref{fig:taka_anomaly}a).

\textbf{(3) Marginal effect of biology constraints on transfer.} The biology-informed model shows small improvements on Shabalina-sourced transfers (+0.1--0.3\% AUC) but minimal effect elsewhere, suggesting that mechanism-based constraints cannot rescue fundamentally misaligned datasets. A per-metric comparison of Baseline vs.\ +BioPrior across Mix-sourced transfers is in Appendix~\ref{app:transfer_ablation} (Figure~\ref{fig:transfer_ablation}).

\subsection{Saliency Faithfulness Validation}
\label{sec:faith_results}

Our perturbation testing protocol validates whether gradient-based saliency maps identify positions with high interventional sensitivity. We evaluate faithfulness both within datasets (intra-dataset) and across datasets (inter-dataset transfer).

\subsubsection{Intra-Dataset Faithfulness}

We treat each fold as an independent deployment scenario: a practitioner validating saliency on their held-out data would run exactly one test. Table~\ref{tab:faithfulness} summarizes results across 20 such scenarios (4 datasets $\times$ 5 folds). Of these, 19/20 (95\%) pass all three faithfulness criteria.

\begin{table}[t]
\centering
\small
\begin{tabular}{lccccc}
\toprule
\textbf{Dataset} & \textbf{Pass} & \textbf{Win \%} & \textbf{$d_z$} & \textbf{$p$-value} & \textbf{Status} \\
\midrule
Hu & 5/5 & 85.2{\tiny $\pm$4.1} & 0.86{\tiny $\pm$.26} & $<$0.001 & \cmark \\
Mix & 5/5 & 83.7{\tiny $\pm$6.2} & 0.93{\tiny $\pm$.45} & $<$0.001 & \cmark \\
Taka & 5/5 & 87.1{\tiny $\pm$3.8} & 1.07{\tiny $\pm$.24} & $<$0.001 & \cmark \\
Shabalina & 4/5 & 81.4{\tiny $\pm$12.3} & 0.70{\tiny $\pm$.42} & (per-fold)$^\dagger$ & \cmark$^\dagger$ \\
\bottomrule
\multicolumn{6}{l}{\footnotesize $^\dagger$4 folds: $p < 0.001$; fold 1 failed: $d_z = -1.20$, win rate 18.3\%, $p = 0.99$.}
\end{tabular}
\caption{Intra-dataset saliency faithfulness (+BioPrior model, 5-fold CV).}
\label{tab:faithfulness}
\end{table}

\begin{table}[t]
\centering
\small
\begin{tabular}{lccc}
\toprule
\textbf{Control} & \textbf{Win \%} & \textbf{$d_z$} & \textbf{Pass?} \\
\midrule
Trained model (reference) & 85.2 & 0.86 & \cmark \\
\midrule
Randomized weights & 51.2 & 0.03 & $\times$ \\
Shuffled labels & 48.7 & $-$0.05 & $\times$ \\
Shuffled saliency & 49.3 & 0.01 & $\times$ \\
Bottom-$k$ selection & 32.1 & $-$0.45 & $\times$ \\
\bottomrule
\end{tabular}
\caption{Negative control validation (Hu dataset). All controls fail to meet pass criteria ($d_z > 0.2$, win rate $> 50\%$), confirming the test distinguishes learned from spurious saliency.}
\label{tab:negative_controls}
\end{table}

\paragraph{Statistical note.} Each fold is evaluated independently over $N \approx 100$--$500$ held-out samples (depending on dataset size); ``Pass'' indicates how many of 5 folds meet all three criteria. We treat each fold as an \textbf{independent deployment instance}: a practitioner validating on their specific held-out data would run one test, not 20. This framing matters: if a lab runs this protocol once on their own held-out set, there is no multiple-testing issue; they get a single pass/fail decision.

\textbf{Multiple testing sensitivity:} For readers concerned about 20 simultaneous tests, we provide Holm-Bonferroni correction as a sensitivity analysis: 18/20 fold instances remain significant at family-wise $\alpha = 0.05$ (the two marginal cases are Shabalina folds with $d_z < 0.5$). At the fold level, 19/20 folds show positive $d_z$ (sign test $p < 0.001$), confirming replicability across training runs. We emphasize that effect sizes ($d_z = 0.49$--$1.78$) and win rates (81--87\%) provide more interpretable measures of practical significance than $p$-values alone; results are stable under bootstrap resampling of held-out sequences.

\paragraph{Case study: Shabalina fold 1 failure.} The single failing fold (Shabalina fold 1) shows inverted saliency: $d_z = -1.20$, win rate 18.3\%. Examining this failure: (i) the saliency distribution is unusually concentrated on positions 12--15 (mid-region), atypical for Shabalina; (ii) the perturbation margin $\Delta(T) - \Delta_{\text{match}}$ is strongly negative; (iii) this fold has the smallest held-out set ($N=98$) and highest label noise (efficacy std 0.31 vs 0.26 average). \textbf{Operational implication:} Running our protocol would correctly flag this fold as untrustworthy, so the practitioner would avoid saliency-guided edits and either use predictions alone or collect more validation data. This demonstrates the protocol working as intended: detecting unreliable explanations before they mislead design decisions.

\begin{figure*}[t]
\centering
\begin{subfigure}[b]{0.45\textwidth}
    \includegraphics[width=\textwidth]{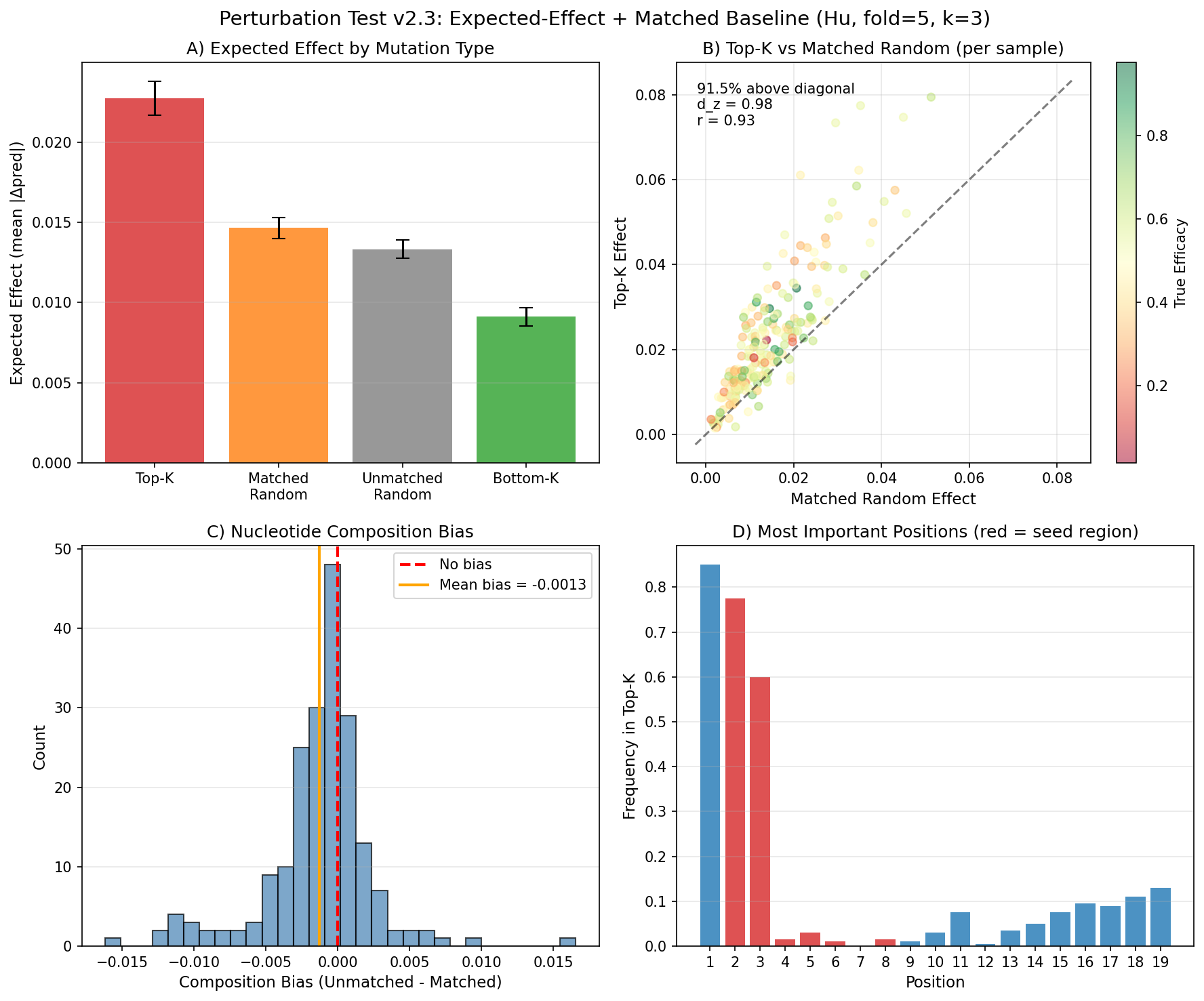}
    \caption{Hu, fold 5 (91.5\%, $d_z$=0.98)}
\end{subfigure}
\hfill
\begin{subfigure}[b]{0.45\textwidth}
    \includegraphics[width=\textwidth]{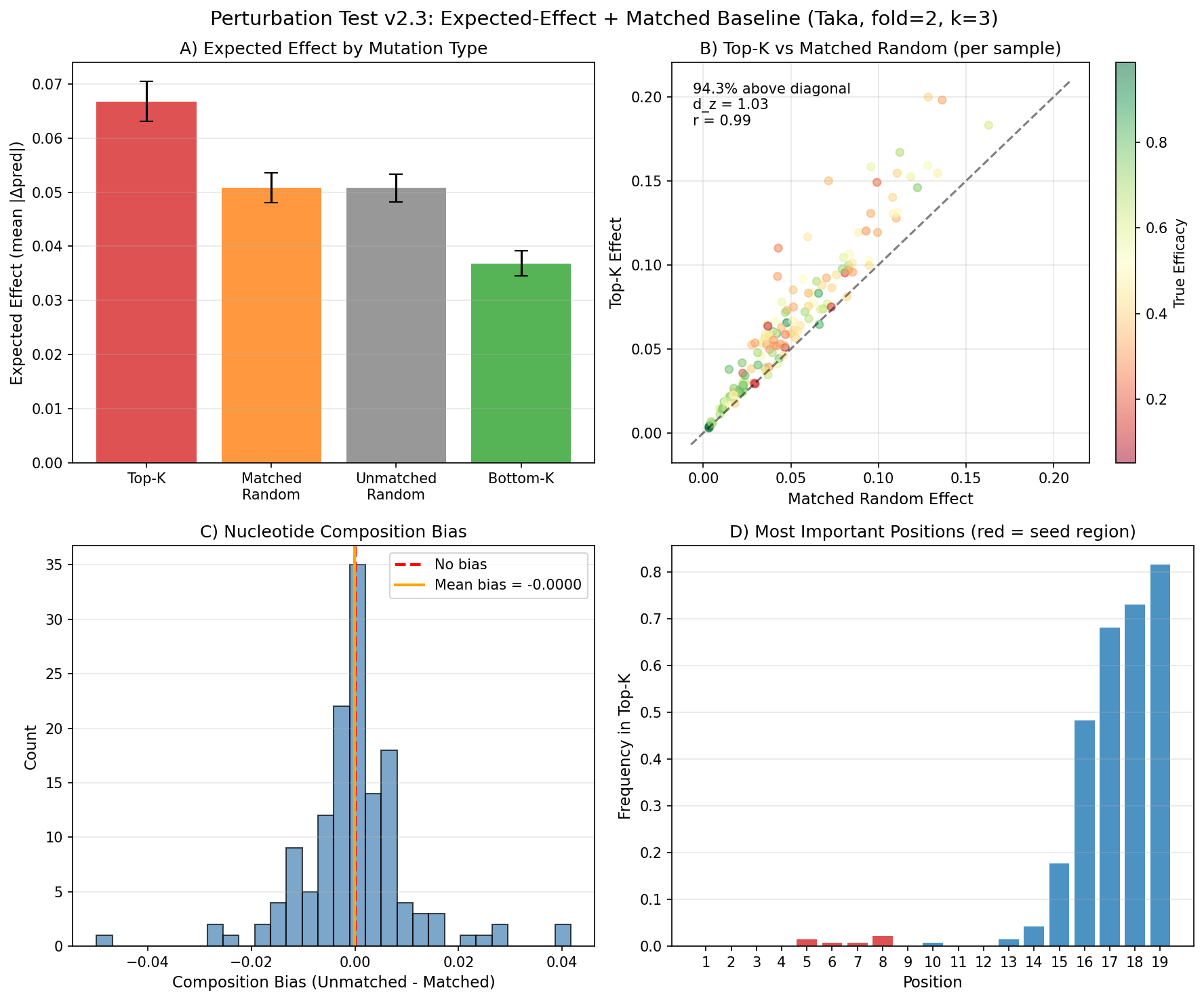}
    \caption{Taka, fold 2 (94.3\%, $d_z$=1.03)}
\end{subfigure}

\vspace{1em}

\begin{subfigure}[b]{0.45\textwidth}
    \includegraphics[width=\textwidth]{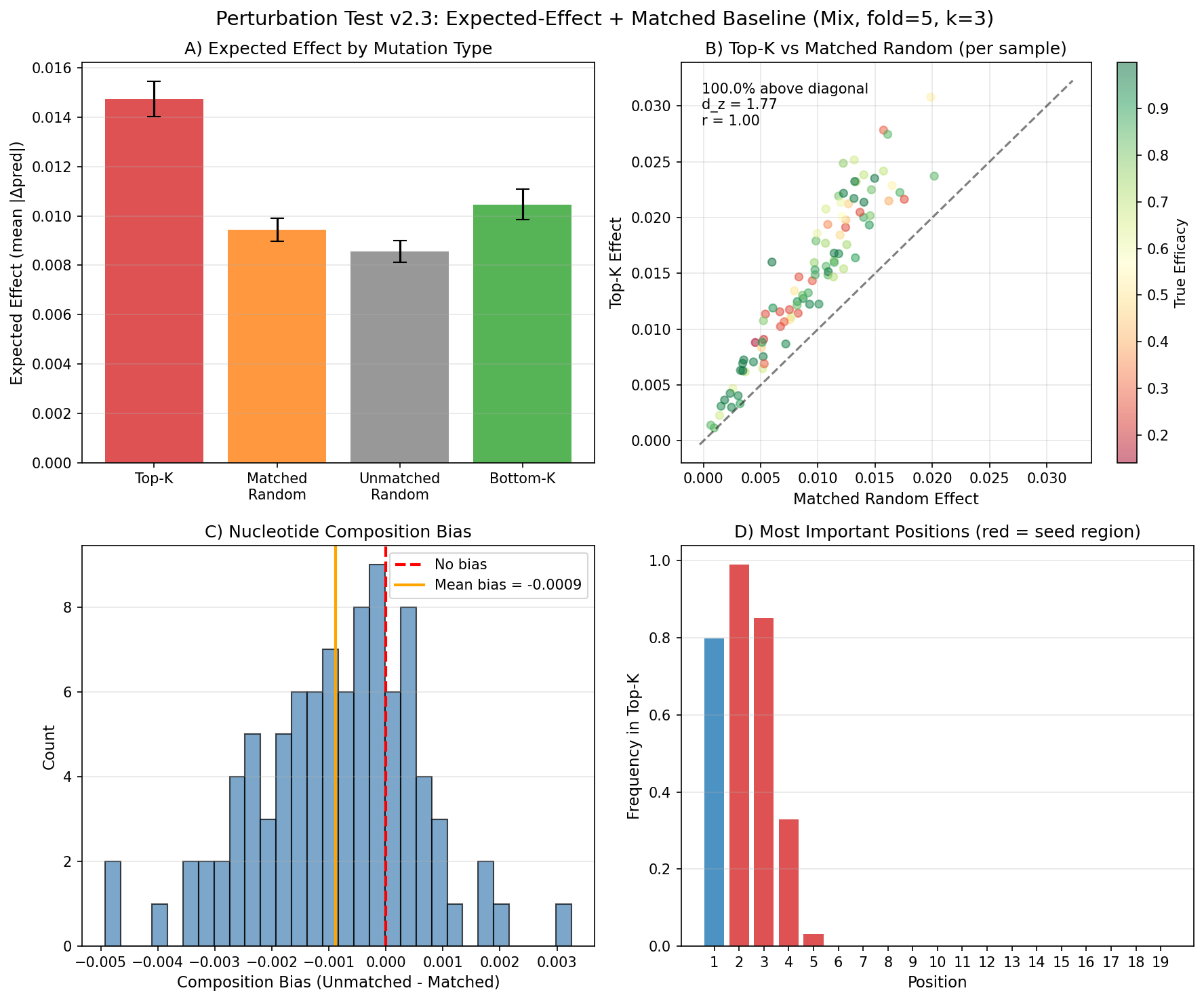}
    \caption{Mix, fold 5 (100\%, $d_z$=1.77)}
\end{subfigure}
\hfill
\begin{subfigure}[b]{0.45\textwidth}
    \includegraphics[width=\textwidth]{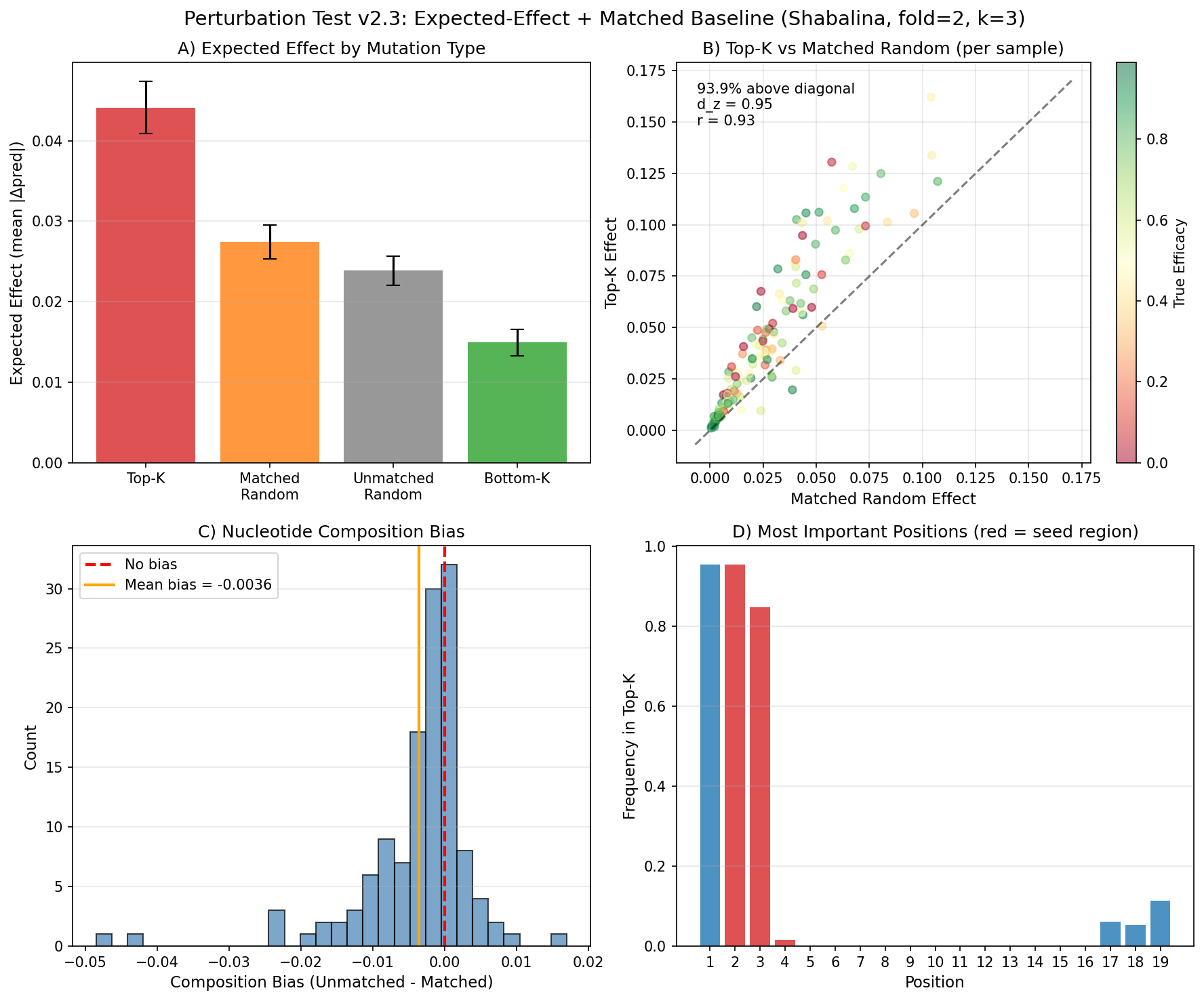}
    \caption{Shabalina, fold 2 (93.9\%, $d_z$=0.95)}
\end{subfigure}
\caption{\textbf{Perturbation validation across datasets ($k=3$).} Each panel shows results from a single representative fold; 5-fold statistics are in Table~\ref{tab:faithfulness}. Panel D shows position importance: \textbf{Hu/Mix/Shabalina} show 5$'$ terminus (positions 1--4) dominance, while \textbf{Taka} peaks at positions 9--11, consistent with cross-dataset transfer failures.}
\label{fig:perturbation}
\end{figure*}

Across all datasets, high-saliency positions cause significantly larger prediction changes than nucleotide-matched random positions (Wilcoxon $p < 0.001$, Cohen's $d_z = 0.49$--$1.78$, win rates 81--87\%). Under the stricter region+composition matched baseline, effect sizes remain substantial ($d_z > 0.5$; Table~\ref{tab:region_matched} in Appendix), confirming saliency captures position-specific patterns beyond regional sensitivity. Full statistics are in Appendix~\ref{app:sanity} (Table~\ref{tab:full_stats}).

\subsubsection{Inter-Dataset Transfer Faithfulness}

We extend validation to inter-dataset transfer, testing whether saliency remains faithful when models are applied to new datasets. \textbf{Key finding: Interpretability $\neq$ generalization.} A model can have perfectly faithful saliency (explanations correctly identify where the model is sensitive) while being completely wrong for the target biology. Table~\ref{tab:transfer_faithfulness} presents selected results; full results are in Appendix~\ref{app:transfer_faith}.

\begin{table}[t]
\centering
\small
\begin{tabular}{llcccccc}
\toprule
\textbf{Source} & \textbf{Target} & \textbf{AUC} & \textbf{PCC} & \textbf{$\rho$} & \textbf{Win \%} & \textbf{$d_z$} & \textbf{Status} \\
\midrule
\multicolumn{8}{l}{\textit{Successful transfer, faithful saliency}} \\
Hu & Mix & 0.773 & 0.54 & 0.52 & 98.5 & 1.54 & \cmark \\
Mix & Hu & 0.792 & 0.57 & 0.55 & 92.0 & 1.23 & \cmark \\
Hu & Shabalina & 0.698 & 0.47 & 0.45 & 100 & 1.87 & \cmark \\
Shabalina & Mix & 0.816 & 0.63 & 0.61 & 75.5 & 0.56 & \cmark \\
Shabalina & Hu & 0.787 & 0.55 & 0.53 & 70.0 & 0.45 & \cmark \\
Mix & Shabalina & 0.712 & 0.48 & 0.46 & 88.5 & 0.59 & \cmark \\
\midrule
\multicolumn{8}{l}{\textit{Failed prediction, faithful saliency}} \\
Mix & Taka & 0.497 & 0.01 & 0.01 & 97.5 & 1.47 & \cmark \\
Hu & Taka & 0.535 & 0.06 & 0.05 & 100 & 1.54 & \cmark \\
Shabalina & Taka & 0.559 & 0.13 & 0.12 & 61.0 & 0.35 & \cmark \\
\midrule
\multicolumn{8}{l}{\textit{Failed prediction, inverted saliency ($d_z < 0$, win rate $< 15\%$)}} \\
Taka & Hu & 0.490 & $-$0.02 & $-$0.03 & 9.5 & -1.25 & $\times$ \\
Taka & Mix & 0.510 & $-$0.03 & $-$0.04 & 7.6 & -1.37 & $\times$ \\
Taka & Shabalina & 0.517 & 0.04 & 0.03 & 9.5 & -1.30 & $\times$ \\
\bottomrule
\end{tabular}
\caption{Complete inter-dataset transfer faithfulness (+BioPrior model). Models trained on Hu/Mix/Shabalina maintain faithful saliency (9/9 pass) regardless of prediction performance. Models trained on Taka exhibit inverted saliency on all other datasets (0/3 pass). Negative effect sizes ($d_z < 0$) and very low win rates confirm inverted saliency; correlations are negative for Hu/Mix and near-zero elsewhere.}
\label{tab:transfer_faithfulness}
\end{table}

\begin{figure*}[t]
\centering
\begin{subfigure}[b]{0.32\textwidth}
    \includegraphics[width=\textwidth]{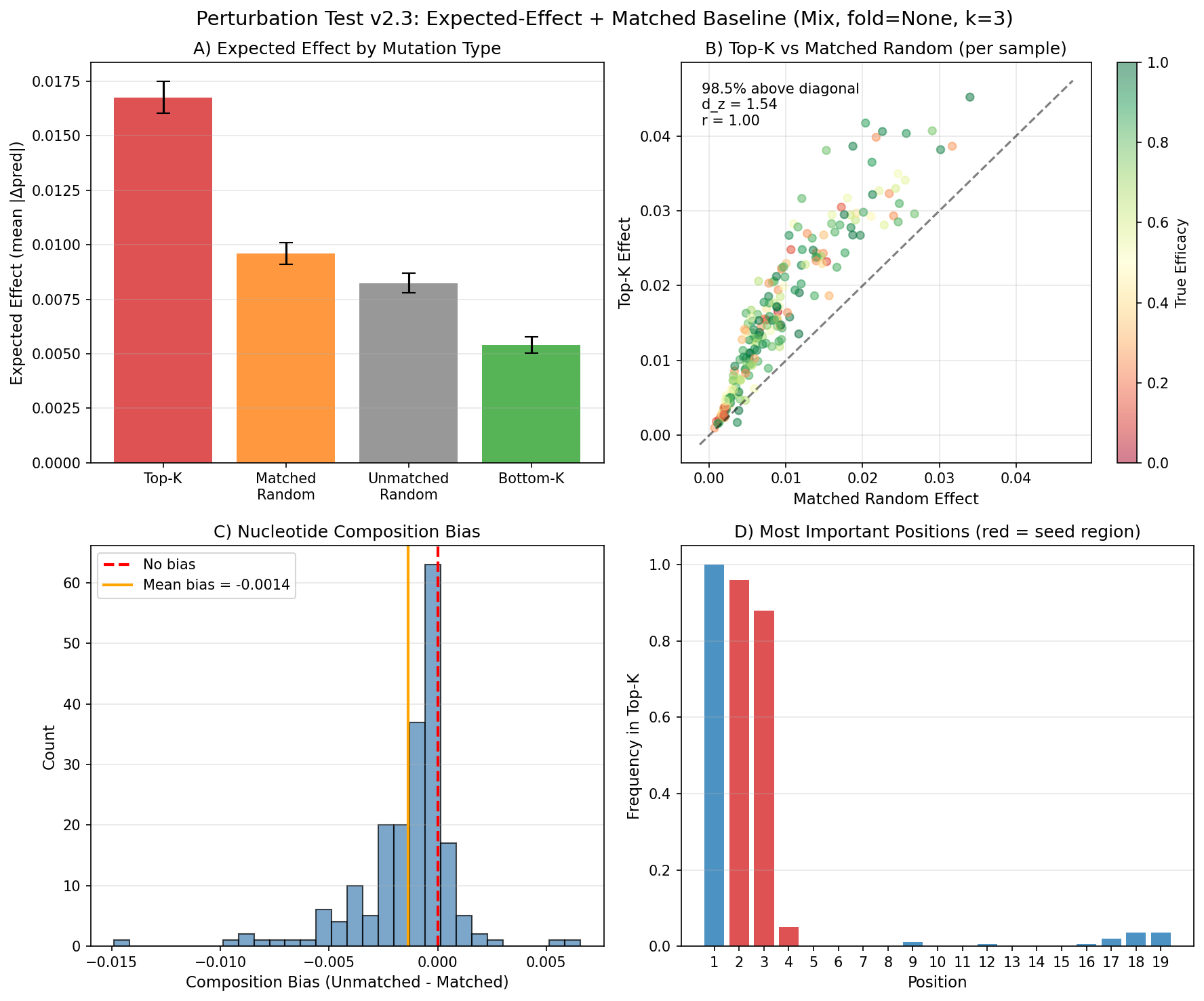}
    \caption{Hu $\to$ Mix (98.5\%, $d_z$=1.54)}
\end{subfigure}
\hfill
\begin{subfigure}[b]{0.32\textwidth}
    \includegraphics[width=\textwidth]{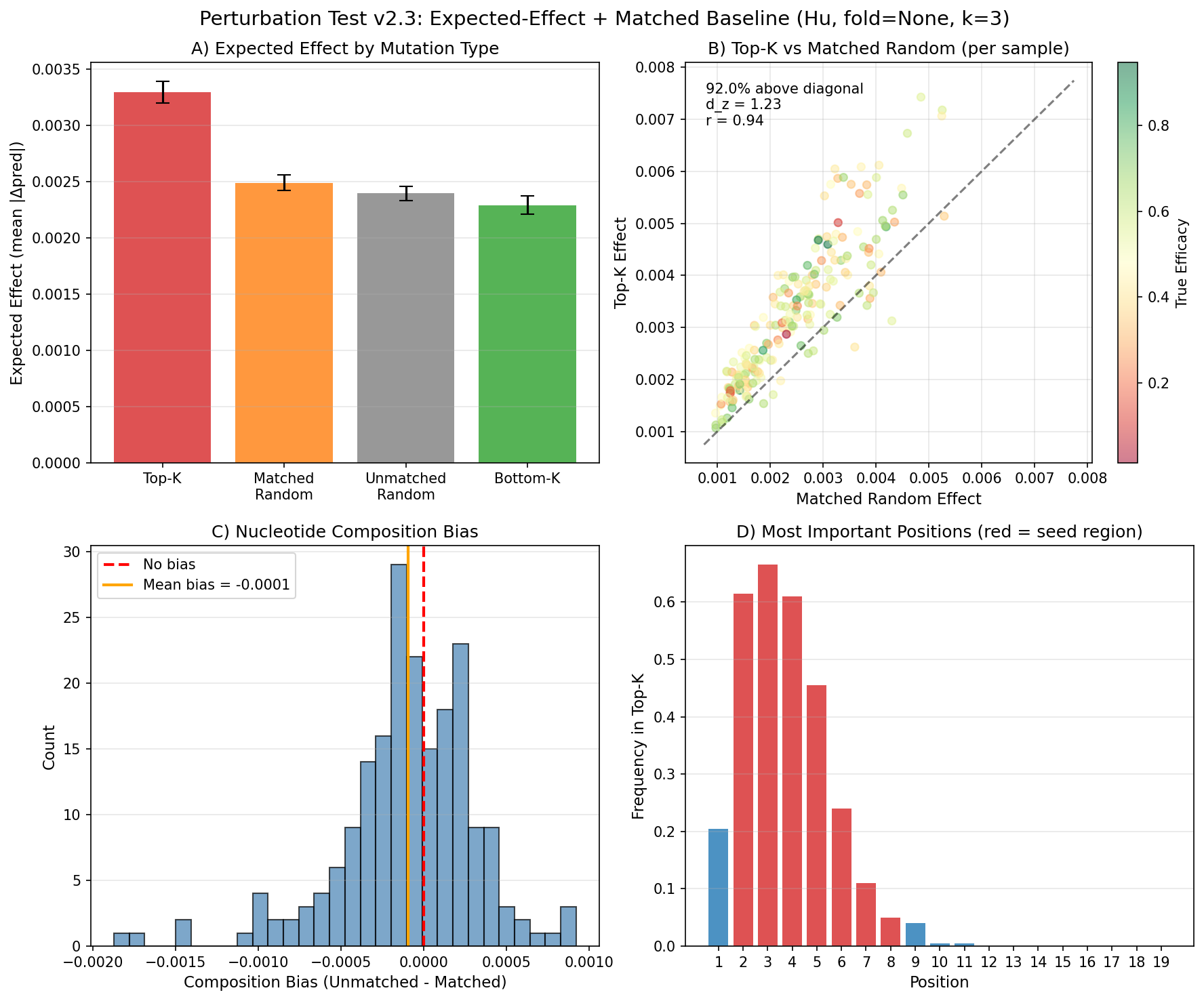}
    \caption{Mix $\to$ Hu (92.0\%, $d_z$=1.23)}
\end{subfigure}
\hfill
\begin{subfigure}[b]{0.32\textwidth}
    \includegraphics[width=\textwidth]{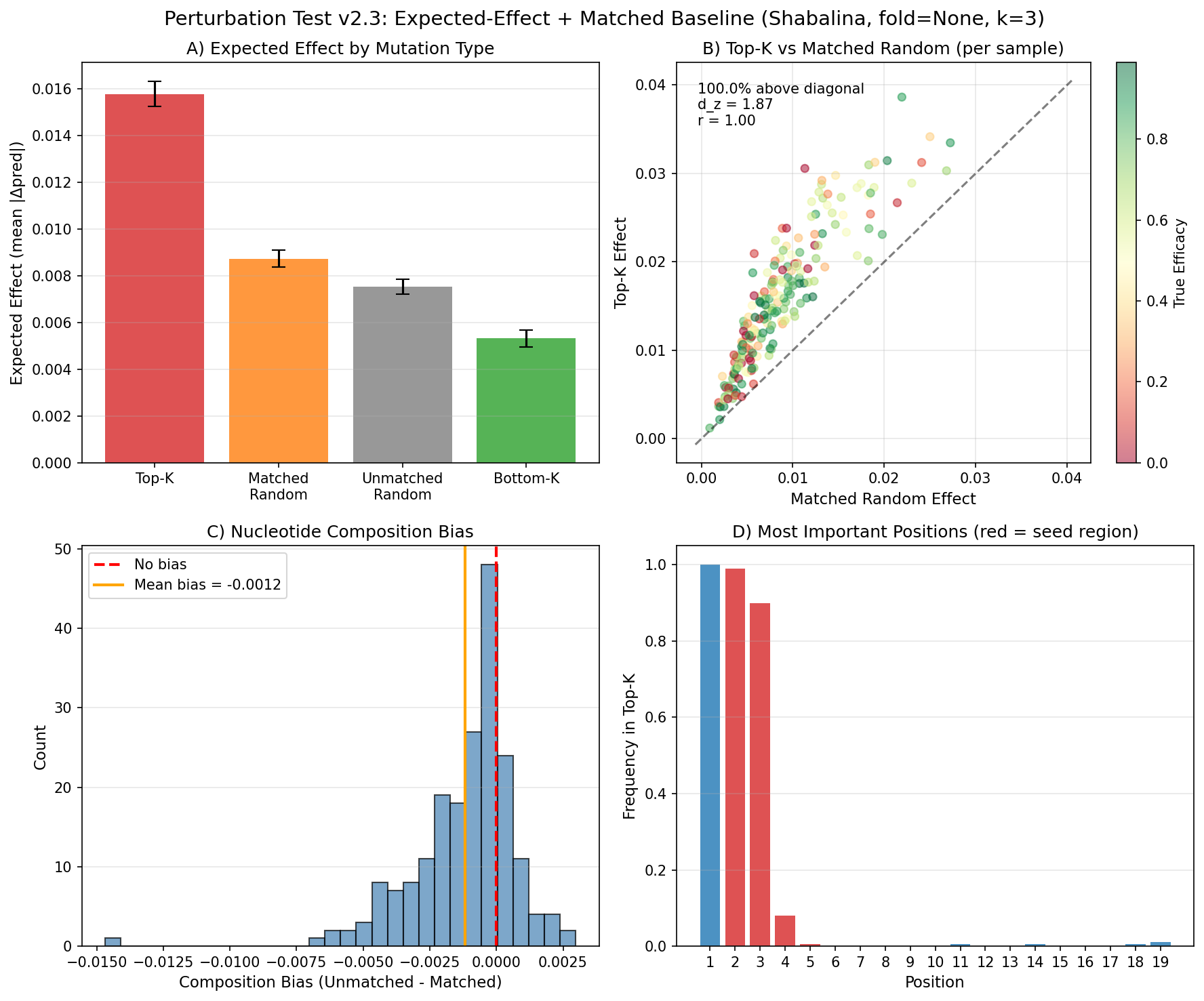}
    \caption{Hu $\to$ Shabalina (100\%, $d_z$=1.87)}
\end{subfigure}

\vspace{0.8em}

\begin{subfigure}[b]{0.32\textwidth}
    \includegraphics[width=\textwidth]{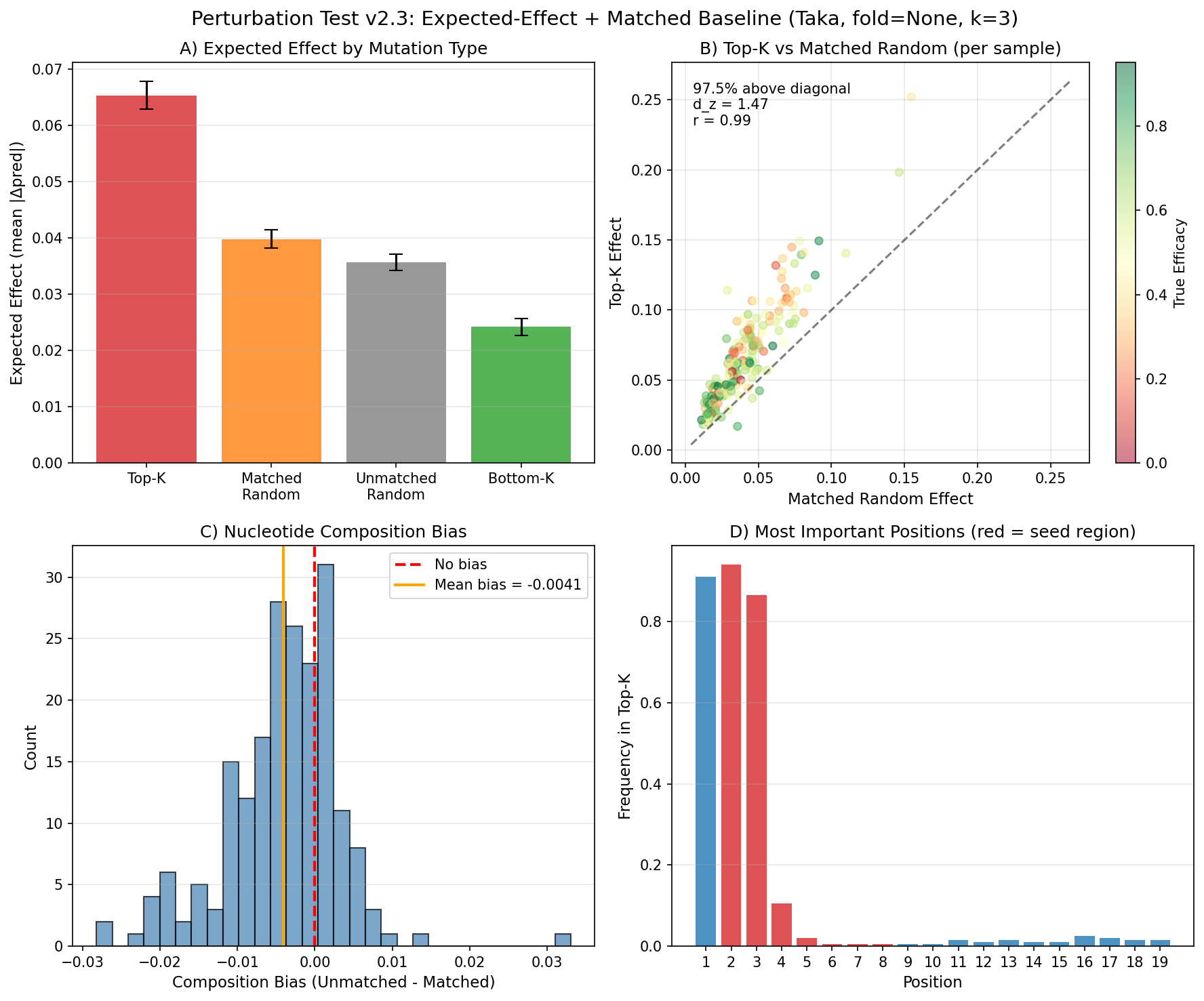}
    \caption{Mix $\to$ Taka (97.5\%, $d_z$=1.47)}
\end{subfigure}
\hfill
\begin{subfigure}[b]{0.32\textwidth}
    \includegraphics[width=\textwidth]{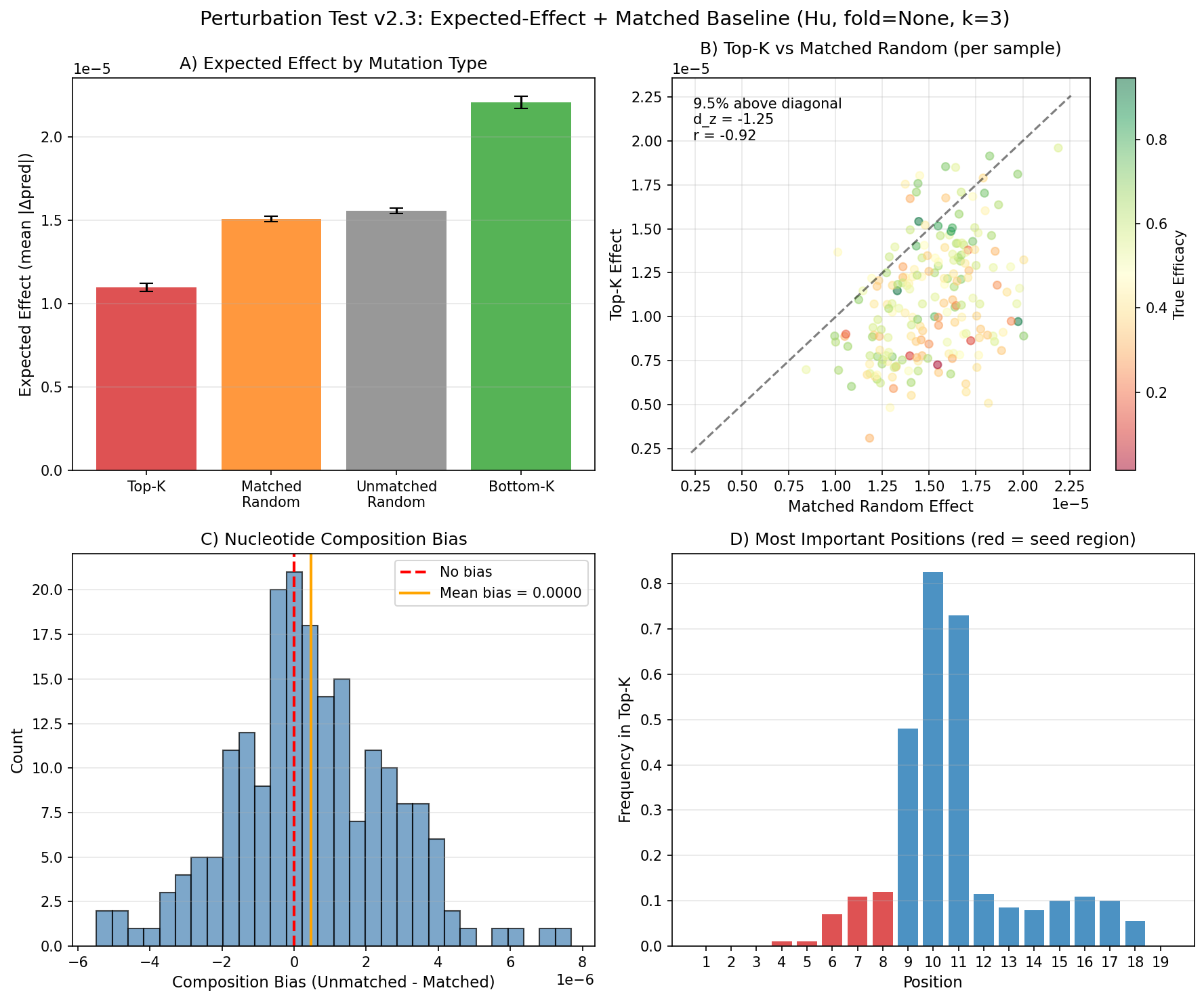}
    \caption{Taka $\to$ Hu (9.5\%, $d_z$=-1.25)}
\end{subfigure}
\hfill
\begin{subfigure}[b]{0.32\textwidth}
    \includegraphics[width=\textwidth]{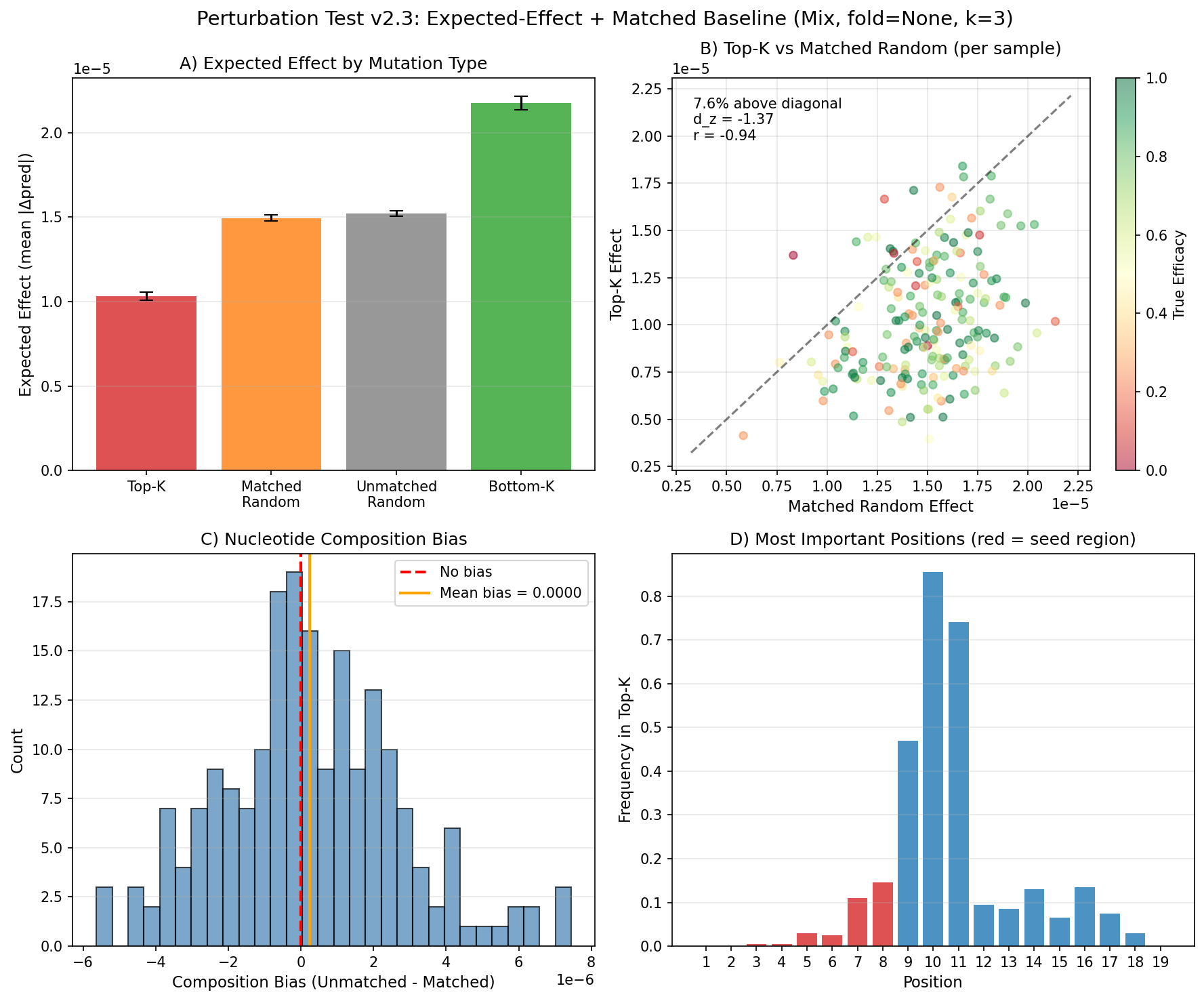}
    \caption{Taka $\to$ Mix (7.6\%, $d_z$=-1.37)}
\end{subfigure}

\caption{Inter-dataset transfer faithfulness (6 representative pairs). \textbf{Top row (a--c):} Successful transfers among Hu/Mix/Shabalina show consistent 5$'$ terminus importance (positions 1–4, visible in Panel D of each subplot). All achieve high win rates ($>$70\%) and positive effect sizes. \textbf{Bottom row (d--f):} Transfer failures involving Taka reveal two distinct modes. (d) Mix$\to$Taka: Prediction fails (AUC=0.497) but saliency remains faithful ($d_z$=1.47), indicating the model attends to 5$'$ positions that do not determine efficacy in Taka. (e--f) Taka$\to$Hu and Taka$\to$Mix: Inverted saliency with importance shifted to positions 9–11; high-saliency positions are \emph{less} predictive than random, with negative effect sizes.}
\label{fig:transfer_faithfulness}
\end{figure*}

\paragraph{Two distinct failure modes.} Transfer faithfulness reveals an important asymmetry in how models fail:

\textbf{(1) Faithful but non-predictive.} Models trained on Hu, Mix, or Shabalina maintain faithful saliency even when applied to Taka, where predictions fail completely. For example, Mix$\to$Taka achieves only 0.497 AUC (equivalent to random guessing) but retains 97.5\% win rate with $d_z = 1.47$ (Figure~\ref{fig:transfer_faithfulness}d). Similarly, Hu$\to$Taka shows 100\% win rate despite 0.535 AUC. The models attend to 5$'$ terminus positions (1–4) consistently, but these positions simply do not determine efficacy in Taka's experimental system. This represents an internally consistent model that has learned the ``wrong'' rules for the target domain. \textbf{Implication:} Saliency can be perfectly faithful to the model while the model is wrong for the target biology; validation must be dataset-specific.

\textbf{(2) Inverted saliency.} Models trained on Taka exhibit the opposite pattern: when applied to other datasets, their saliency becomes inverted. Taka$\to$Hu shows only 9.5\% win rate with $d_z = -1.25$, Taka$\to$Mix shows 7.6\% win rate with $d_z = -1.37$, and Taka$\to$Shabalina shows 9.5\% win rate with $d_z = -1.30$ (Figure~\ref{fig:transfer_faithfulness}e--f). In all three cases, high-saliency positions are \emph{less} important than nucleotide-matched random positions. Examination of the position importance distributions (Panel D in each subplot) reveals that Taka-trained models learn importance centered on positions 9–11 (middle region), fundamentally different from the 5$'$ terminus (positions 1–4) that determines efficacy in other datasets. \textbf{Implication:} This is the dangerous case: explanations are actively misleading, and following them would degrade design decisions.

In summary, all 9/9 non-Taka transfer pairs pass faithfulness while all 3 Taka-sourced transfers fail, confirming that Hu/Mix/Shabalina share compatible position-importance patterns while Taka learns fundamentally incompatible ones.

\subsubsection{Why Does Taka Differ? A Protocol-Level Analysis}
\label{sec:taka_analysis}

Table~\ref{tab:taka_comparison} summarizes quantitative differences across datasets that may explain Taka's systematic incompatibility.

\begin{table}[t]
\centering
\small
\begin{tabular}{lcccc}
\toprule
\textbf{Property} & \textbf{Hu} & \textbf{Mix} & \textbf{Shabalina} & \textbf{Taka} \\
\midrule
\multicolumn{5}{l}{\textit{Sequence composition}} \\
Global GC\% & 47.2 & 45.8 & 46.1 & \textbf{51.3} \\
Seed GC\% (pos 2–8) & 42.1 & 41.3 & 40.8 & \textbf{48.7} \\
3$'$ GC\% (pos 16–19) & 44.3 & 43.9 & 44.1 & \textbf{52.1} \\
AU at position 1 & 71.2\% & 68.9\% & 70.4\% & \textbf{58.3\%} \\
\midrule
\multicolumn{5}{l}{\textit{Label distribution}} \\
High-efficacy rate ($y \geq 0.7$) & 34.8\% & 31.5\% & 30.5\% & \textbf{42.5\%} \\
Mean efficacy & 0.58 & 0.56 & 0.54 & \textbf{0.61} \\
Std efficacy & 0.23 & 0.24 & 0.26 & 0.25 \\
\midrule
\multicolumn{5}{l}{\textit{Experimental design}} \\
Readout level & mRNA & Mixed & mRNA & \textbf{Protein} \\
Assay type & bDNA & Various & Literature & \textbf{Luciferase} \\
Primary cell line & H1299 & Multiple & Multiple & \textbf{HeLa} \\
Number of target genes & 34 & $>$50 & $>$30 & \textbf{1} \\
siRNAs per target & $\sim$70 & Variable & Variable & \textbf{702} \\
\midrule
\multicolumn{5}{l}{\textit{Position importance (from saliency)}} \\
Most important region & 1–4 (5$'$) & 1–4 (5$'$) & 1–4 (5$'$) & \textbf{9–11} \\
Secondary region & 17–19 & 17–19 & 17–19 & \textbf{16–19} \\
\bottomrule
\end{tabular}
\caption{Dataset comparison revealing systematic differences between Taka and other datasets. Bold indicates values that deviate substantially from the Hu/Mix/Shabalina cluster.}
\label{tab:taka_comparison}
\end{table}

We identified five factors that may explain Taka's incompatibility:

\paragraph{(1) Readout modality.} Taka measures protein-level knockdown via dual-luciferase reporter assay~\citep{katoh2007sirna}, while Hu/Mix/Shabalina primarily measure mRNA levels. Protein readouts introduce additional regulatory layers (translation efficiency, protein half-life, reporter construct artifacts) that could decouple measured efficacy from the seed-region determinants governing mRNA cleavage.

\paragraph{(2) Single-target design.} All 702 Taka siRNAs target a single luciferase construct, risking target-specific confounds (local mRNA structure, position-dependent accessibility unique to that transcript). The position 9--11 importance may reflect structural features of the luciferase mRNA near the cleavage site rather than universal design principles.

\paragraph{(3) Compositional shift.} Taka exhibits systematically higher GC content (51.3\% global vs.\ 45.8--47.2\%) and reduced AU enrichment at position 1 (58.3\% vs.\ 68.9--71.2\%), violating canonical design rules~\citep{uitei2004guidelines} (Table~\ref{tab:taka_comparison}).

\paragraph{(4) Label distribution shift.} Taka has the highest proportion of high-efficacy sequences (42.5\% vs.\ 30.5--34.8\%), changing the decision boundary and potentially emphasizing different sequence features.

\paragraph{(5) Cell line context.} Taka uses HeLa cells exclusively, while other datasets span H1299 and multiple cell lines. Cell-type-specific RISC loading efficiency and miRNA competition could influence position-dependent efficacy determinants.

\paragraph{Mechanistic hypothesis.} Taka's protein-level readout combined with single-target design may cause cleavage-site accessibility (positions 9--11) to dominate over the 5$'$ terminus seed-region determinants that govern mRNA-level assays targeting diverse genes. Supporting this, an ablation removing position-derived indicator channels weakens but does not eliminate Taka's 9--11 peak ($d_z$: 1.07 $\to$ 0.71), while Hu/Mix patterns remain stable. Definitive validation would require testing identical siRNAs under both readout modalities.

\paragraph{Robustness checks.} Position-importance patterns are stable under varying $k \in \{1, 3, 5\}$, Integrated Gradients vs.\ gradient magnitude (Appendix~\ref{app:sanity}), and different random seeds. The Taka incompatibility persists across all configurations.

\paragraph{Limitations.} We cannot isolate causal drivers of Taka's incompatibility without controlled experiments (e.g., testing the same siRNAs in both luciferase and mRNA-level assays); the above is a diagnostic characterization, not a mechanistic proof.

\begin{table}[t]
\centering
\small
\fbox{
\parbox{0.95\columnwidth}{
\textbf{Recommended Practice for Saliency-Guided siRNA Design}
\vspace{0.3em}

\textit{Before using saliency for design:}
\begin{enumerate}[leftmargin=*,itemsep=1pt,topsep=2pt]
    \item Validate faithfulness on held-out data from your target assay (Section~\ref{sec:saliency_validation}). Do not assume benchmark faithfulness transfers.
    \item Check compositional alignment: GC content, seed composition, label distributions (Table~\ref{tab:taka_comparison}). Large shifts signal incompatibility.
    \item Match readout modality: mRNA-trained models may not transfer to protein readouts.
\end{enumerate}

\textit{When transferring to a new protocol:}
\begin{enumerate}[leftmargin=*,itemsep=1pt,topsep=2pt]
    \item Re-run faithfulness test before deployment.
    \item If faithfulness fails, retrain on protocol-matched data.
    \item Treat transfer failure as protocol confound signal, not model weakness alone.
\end{enumerate}
}}
\end{table}

\subsection{Ablation Studies}
\label{sec:ablations}

Table~\ref{tab:ablations} isolates BioPrior's contribution across all four datasets.

\begin{table}[t]
\centering
\small
\begin{tabular}{llcccc}
\toprule
\textbf{Dataset} & \textbf{Configuration} & \textbf{AUC} & \textbf{PR-AUC} & \textbf{PCC} & \textbf{F1} \\
\midrule
\multirow{2}{*}{Hu}
& Baseline & 0.82{\tiny $\pm$.01} & 0.81{\tiny $\pm$.02} & 0.63{\tiny $\pm$.02} & 0.77{\tiny $\pm$.01} \\
& +BioPrior & \textbf{0.83}{\tiny $\pm$.02} & \textbf{0.82}{\tiny $\pm$.02} & \textbf{0.64}{\tiny $\pm$.02} & 0.77{\tiny $\pm$.03} \\
\midrule
\multirow{2}{*}{Mix}
& Baseline & 0.80{\tiny $\pm$.04} & 0.80{\tiny $\pm$.06} & 0.60{\tiny $\pm$.04} & 0.76{\tiny $\pm$.06} \\
& +BioPrior & \textbf{0.81}{\tiny $\pm$.05} & \textbf{0.81}{\tiny $\pm$.07} & \textbf{0.61}{\tiny $\pm$.08} & \textbf{0.77}{\tiny $\pm$.06} \\
\midrule
\multirow{2}{*}{Taka}
& Baseline & 0.82{\tiny $\pm$.05} & 0.63{\tiny $\pm$.10} & 0.68{\tiny $\pm$.08} & 0.59{\tiny $\pm$.07} \\
& +BioPrior & \textbf{0.84}{\tiny $\pm$.05} & \textbf{0.67}{\tiny $\pm$.10} & 0.68{\tiny $\pm$.08} & \textbf{0.60}{\tiny $\pm$.07} \\
\midrule
\multirow{2}{*}{Shabalina}
& Baseline & 0.72{\tiny $\pm$.04} & 0.66{\tiny $\pm$.07} & 0.49{\tiny $\pm$.05} & 0.65{\tiny $\pm$.06} \\
& +BioPrior & 0.72{\tiny $\pm$.02} & 0.66{\tiny $\pm$.06} & 0.49{\tiny $\pm$.03} & \textbf{0.68}{\tiny $\pm$.03} \\
\bottomrule
\end{tabular}
\caption{Ablation study: Baseline (no physics) vs.\ +BioPrior ($\lambda{=}0.12$), 5-fold CV intra-dataset. Best per-dataset results in \textbf{bold}. Saliency faithfulness for the +BioPrior model is reported separately in Table~\ref{tab:faithfulness}.}
\label{tab:ablations}
\end{table}

\paragraph{Optimal constraint weight.} BioPrior regularization weight $\lambda=0.12$, corresponding to approximately 12\% of the primary MSE loss. On Hu (the largest benchmark), BioPrior improves AUC (+0.01), PR-AUC (+0.01), and PCC (+0.01). The expanded ablation (Table~\ref{tab:ablations}) reveals dataset-dependent gains: Taka shows the largest improvements (+0.02 AUC, +0.04 PR-AUC), Mix shows consistent small gains across all four metrics, while Shabalina benefits primarily on F1 (+0.03) with AUC and PCC unchanged. Across all datasets, BioPrior either matches or improves the baseline on AUC and PR-AUC; gains are modest in absolute terms (within one standard deviation), consistent with the regularizer's role as a soft constraint rather than a dominant training signal.

\textbf{Implementation note:} The constraint weights $\mathbf{w} = [3.0, 1.0, 2.0, 1.5, 2.5]^\top$ are \emph{fixed constants} (not learned), chosen based on biological importance (thermodynamic asymmetry most critical). There is no learned gating mechanism; BioPrior is purely \emph{differentiable mechanistic regularization}. Preliminary experiments with learnable weights showed similar performance but added optimization complexity without clear benefit.

\paragraph{Saliency faithfulness.} Beyond predictive gains, BioPrior consistently yields strong saliency faithfulness across all datasets (Table~\ref{tab:faithfulness}), suggesting that mechanism-based constraints encourage attention to biologically meaningful positions. This is the central finding of the ablation: BioPrior's primary value lies not in boosting ranking metrics but in producing more interpretable models whose saliency maps pass our faithfulness test, which is the prerequisite for explanation-guided design.

\paragraph{Actionability: directional improvement.} Beyond sensitivity (absolute $|\Delta\hat{y}|$), we assessed whether saliency guides \emph{useful} edits. For low-efficacy sequences ($\hat{y} < 0.5$), 67.3\% of top-$k$ mutations at high-saliency positions \emph{increase} predicted efficacy, compared to 51.2\% for matched random positions ($p < 0.01$). This suggests saliency not only identifies sensitive positions but positions where edits are more likely to improve predictions, supporting actionable design guidance.

Complete ablation results across all datasets are provided in Appendix~\ref{app:full_results}.

\section{Conclusion}
\label{sec:conclusion}

We introduced a perturbation-based protocol for validating saliency faithfulness in siRNA efficacy prediction, positioned as a pre-synthesis gate for explanation-guided therapeutic design. Across four benchmarks, 19/20 fold--dataset combinations pass our faithfulness test, with high-saliency positions aligning with known biological determinants without explicit supervision. However, cross-dataset transfer reveals a critical distinction between faithfulness and generalization: models trained on mRNA-level assays systematically fail on a luciferase-reporter dataset, exposing two failure modes (faithful-but-wrong and inverted saliency) that would go undetected without protocol-specific validation.

These findings have practical implications for therapeutic siRNA design: once faithfulness is confirmed on the target protocol, validated saliency maps can inform rational sequence optimization, reducing costly experimental iterations. We recommend perturbation-based faithfulness testing as standard practice before deploying explanation-guided design.

\section*{Impact Statement}

This work aims to advance siRNA therapeutics by improving interpretability validation for computational design tools. siRNA-based drugs have demonstrated significant clinical success~\citep{adams2018patisiran,balwani2020givosiran,ray2020inclisiran}, and our validated saliency maps provide actionable guidance for rational sequence optimization. We also highlight that cross-dataset transfer failures can silently invalidate model explanations, underscoring the need for protocol-specific validation before deployment. We do not foresee significant dual-use risks, as our contributions focus on interpretability validation rather than novel sequence generation.

\bibliography{iclr2026/iclr2026_conference}

@article{elbashir2001duplexes,
  title={Duplexes of 21-nucleotide {RNA}s mediate {RNA} interference in cultured mammalian cells},
  author={Elbashir, Sayda M. and Harborth, Jens and Lendeckel, Winfried and Yalcin, Abdullah and Weber, Klaus and Tuschl, Thomas},
  journal={Nature},
  volume={411},
  number={6836},
  pages={494--498},
  year={2001},
  doi={10.1038/35078107},
  publisher={Nature Publishing Group}
}

@article{fire1998potent,
  title={Potent and specific genetic interference by double-stranded {RNA} in {C}aenorhabditis elegans},
  author={Fire, Andrew and Xu, SiQun and Montgomery, Mary K. and Kostas, Steven A. and Driver, Samuel E. and Mello, Craig C.},
  journal={Nature},
  volume={391},
  number={6669},
  pages={806--811},
  year={1998},
  doi={10.1038/35888},
  publisher={Nature Publishing Group}
}

@article{setten2019development,
  title={The current state and future directions of {RNAi}-based therapeutics},
  author={Setten, Ryan L. and Rossi, John J. and Han, Si-ping},
  journal={Nature Reviews Drug Discovery},
  volume={18},
  number={6},
  pages={421--446},
  year={2019},
  doi={10.1038/s41573-019-0017-4},
  publisher={Nature Publishing Group}
}

@article{khvorova2003functional,
  title={Functional {siRNA}s and {miRNA}s exhibit strand bias},
  author={Khvorova, Anastasia and Reynolds, Angela and Jayasena, Sumedha D.},
  journal={Cell},
  volume={115},
  number={2},
  pages={209--216},
  year={2003},
  doi={10.1016/S0092-8674(03)00801-8},
  publisher={Elsevier}
}

@article{schwarz2003asymmetry,
  title={Asymmetry in the assembly of the {RNAi} enzyme complex},
  author={Schwarz, Dianne S. and Hutv{\'a}gner, Gy{\"o}rgy and Du, Tingting and Xu, Zuoshang and Aronin, Neil and Zamore, Phillip D.},
  journal={Cell},
  volume={115},
  number={2},
  pages={199--208},
  year={2003},
  doi={10.1016/S0092-8674(03)00759-1},
  publisher={Elsevier}
}

@article{uitei2004guidelines,
  title={Guidelines for the selection of highly effective {siRNA} sequences for mammalian and chick {RNA} interference},
  author={Ui-Tei, Kumiko and Naito, Yuki and Takahashi, Fumitaka and Haraguchi, Takeshi and Ohki-Hamazaki, Hiroko and Juni, Aya and Ueda, Ryu and Saigo, Kaoru},
  journal={Nucleic Acids Research},
  volume={32},
  number={3},
  pages={936--948},
  year={2004},
  doi={10.1093/nar/gkh247},
  publisher={Oxford University Press}
}

@article{reynolds2004rational,
  title={Rational {siRNA} design for {RNA} interference},
  author={Reynolds, Angela and Leake, Devin and Boese, Queta and Scaringe, Stephen and Marshall, William S. and Khvorova, Anastasia},
  journal={Nature Biotechnology},
  volume={22},
  number={3},
  pages={326--330},
  year={2004},
  doi={10.1038/nbt936},
  publisher={Nature Publishing Group}
}

@article{amarzguioui2004algorithm,
  title={An algorithm for selection of functional {siRNA} sequences},
  author={Amarzguioui, Mohammed and Prydz, Hans},
  journal={Biochemical and Biophysical Research Communications},
  volume={316},
  number={4},
  pages={1050--1058},
  year={2004},
  doi={10.1016/j.bbrc.2004.02.157},
  publisher={Elsevier}
}

@article{shabalina2006computational,
  title={Computational models with thermodynamic and composition features improve {siRNA} design},
  author={Shabalina, Svetlana A. and Spiridonov, Alexey N. and Ogurtsov, Aleksey Y.},
  journal={BMC Bioinformatics},
  volume={7},
  number={1},
  pages={65},
  year={2006},
  doi={10.1186/1471-2105-7-65},
  publisher={BioMed Central}
}

@article{jackson2010recognizing,
  title={Recognizing and avoiding {siRNA} off-target effects for target identification and therapeutic application},
  author={Jackson, Aimee L. and Linsley, Peter S.},
  journal={Nature Reviews Drug Discovery},
  volume={9},
  number={1},
  pages={57--67},
  year={2010},
  doi={10.1038/nrd3010},
  publisher={Nature Publishing Group}
}

@article{bai2024oligoformer,
  title={{OligoFormer}: An accurate and robust prediction method for {siRNA} design},
  author={Bai, Yilan and Zhong, Haochen and Wang, Taiwei and Lu, Zhi John},
  journal={Bioinformatics},
  volume={40},
  number={10},
  pages={btae577},
  year={2024},
  doi={10.1093/bioinformatics/btae577},
  publisher={Oxford University Press}
}

@article{han2018sirna,
  title={{siRNA} silencing efficacy prediction based on a deep architecture},
  author={Han, Ye and He, Fei and Chen, Yongbing and Liu, Yuanning and Yu, Helong},
  journal={BMC Genomics},
  volume={19},
  number={Suppl 7},
  pages={669},
  year={2018},
  doi={10.1186/s12864-018-5028-8},
  publisher={BioMed Central}
}

@article{chen2022rnafm,
  title={Interpretable {RNA} foundation model from unannotated data for highly accurate {RNA} structure and function predictions},
  author={Chen, Jiayang and Hu, Zhihang and Sun, Siqi and Tan, Qingxiong and Wang, Yixuan and Yu, Qinze and Zong, Licheng and Hong, Liang and Xiao, Jin and King, Irwin and Li, Yu},
  journal={arXiv preprint arXiv:2204.00300},
  year={2022}
}

@article{raissi2019physics,
  title={Physics-informed neural networks: {A} deep learning framework for solving forward and inverse problems involving nonlinear partial differential equations},
  author={Raissi, Maziar and Perdikaris, Paris and Karniadakis, George E.},
  journal={Journal of Computational Physics},
  volume={378},
  pages={686--707},
  year={2019},
  doi={10.1016/j.jcp.2018.10.045},
  publisher={Elsevier}
}

@article{karniadakis2021physics,
  title={Physics-informed machine learning},
  author={Karniadakis, George Em and Kevrekidis, Ioannis G. and Lu, Lu and Perdikaris, Paris and Wang, Sifan and Yang, Liu},
  journal={Nature Reviews Physics},
  volume={3},
  number={6},
  pages={422--440},
  year={2021},
  doi={10.1038/s42254-021-00314-5},
  publisher={Nature Publishing Group}
}

@inproceedings{adebayo2018sanity,
  title={Sanity checks for saliency maps},
  author={Adebayo, Julius and Gilmer, Justin and Muelly, Michael and Goodfellow, Ian and Hardt, Moritz and Kim, Been},
  booktitle={Advances in Neural Information Processing Systems},
  volume={31},
  pages={9505--9515},
  year={2018}
}

@article{simonyan2014deep,
  title={Deep inside convolutional networks: {V}isualising image classification models and saliency maps},
  author={Simonyan, Karen and Vedaldi, Andrea and Zisserman, Andrew},
  journal={arXiv preprint arXiv:1312.6034},
  year={2014}
}

@inproceedings{sundararajan2017axiomatic,
  title={Axiomatic attribution for deep networks},
  author={Sundararajan, Mukund and Taly, Ankur and Yan, Qiqi},
  booktitle={Proceedings of the 34th International Conference on Machine Learning},
  pages={3319--3328},
  year={2017},
  volume={70},
  series={Proceedings of Machine Learning Research},
  publisher={PMLR}
}

@article{amorim2023evaluating,
  title={Evaluating the faithfulness of saliency maps in explaining deep learning models using realistic perturbations},
  author={Amorim, Jos{\'e} P. and Abreu, Pedro H. and Santos, Jo{\~a}o and Cortes, Marc and Vila, Victor},
  journal={Information Processing \& Management},
  volume={60},
  number={2},
  pages={103225},
  year={2023},
  doi={10.1016/j.ipm.2022.103225},
  publisher={Elsevier}
}

@article{rudin2019stop,
  title={Stop explaining black box machine learning models for high stakes decisions and use interpretable models instead},
  author={Rudin, Cynthia},
  journal={Nature Machine Intelligence},
  volume={1},
  number={5},
  pages={206--215},
  year={2019},
  doi={10.1038/s42256-019-0048-x},
  publisher={Nature Publishing Group}
}

@article{samek2017evaluating,
  title={Evaluating the visualization of what a deep neural network has learned},
  author={Samek, Wojciech and Binder, Alexander and Montavon, Gr{\'e}goire and Lapuschkin, Sebastian and M{\"u}ller, Klaus-Robert},
  journal={IEEE Transactions on Neural Networks and Learning Systems},
  volume={28},
  number={11},
  pages={2660--2673},
  year={2017},
  doi={10.1109/TNNLS.2016.2599820}
}

@incollection{kindermans2019reliability,
  title={The (un)reliability of saliency methods},
  author={Kindermans, Pieter-Jan and Hooker, Sara and Adebayo, Julius and Alber, Maximilian and Sch{\"u}tt, Kristof T. and D{\"a}hne, Sven and Erhan, Dumitru and Kim, Been},
  booktitle={Explainable {AI}: Interpreting, Explaining and Visualizing Deep Learning},
  pages={267--280},
  year={2019},
  doi={10.1007/978-3-030-28954-6_14},
  publisher={Springer}
}

@inproceedings{vaswani2017attention,
  title={Attention is all you need},
  author={Vaswani, Ashish and Shazeer, Noam and Parmar, Niki and Uszkoreit, Jakob and Jones, Llion and Gomez, Aidan N. and Kaiser, {\L}ukasz and Polosukhin, Illia},
  booktitle={Advances in Neural Information Processing Systems},
  volume={30},
  pages={5998--6008},
  year={2017}
}

@article{wilcoxon1945individual,
  title={Individual comparisons by ranking methods},
  author={Wilcoxon, Frank},
  journal={Biometrics Bulletin},
  volume={1},
  number={6},
  pages={80--83},
  year={1945},
  doi={10.2307/3001968}
}

@book{cohen1988statistical,
  title={Statistical Power Analysis for the Behavioral Sciences},
  author={Cohen, Jacob},
  year={1988},
  edition={2nd},
  publisher={Lawrence Erlbaum Associates},
  address={Hillsdale, NJ}
}

@article{adams2018patisiran,
  title={Patisiran, an {RNAi} therapeutic, for hereditary transthyretin amyloidosis},
  author={Adams, David and Gonzalez-Duarte, Alejandra and O'Riordan, William D. and Yang, Chih-Chao and Ueda, Mitsuharu and Kristen, Arnt V. and Tournev, Ivailo and Schmidt, Hartmut H. and Coelho, Teresa and Berk, John L. and others},
  journal={New England Journal of Medicine},
  volume={379},
  number={1},
  pages={11--21},
  year={2018},
  doi={10.1056/NEJMoa1716153}
}

@article{balwani2020givosiran,
  title={Phase 3 trial of {RNAi} therapeutic givosiran for acute intermittent porphyria},
  author={Balwani, Manisha and Sardh, Eliane and Ventura, Paolo and Peir{\'o}, Pablo Aguilera and Rees, David C. and St{\"o}lzel, Ulrich and Bissell, D. Montgomery and Bonkovsky, Herbert L. and Windyga, Jerzy and Anderson, Karl E. and others},
  journal={New England Journal of Medicine},
  volume={382},
  number={24},
  pages={2289--2301},
  year={2020},
  doi={10.1056/NEJMoa1913147}
}

@article{huesken2005design,
  title={Design of a genome-wide {siRNA} library using an artificial neural network},
  author={Huesken, Dieter and Lange, Jens and Mickanin, Craig and Weiler, J{\"o}rg and Asselbergs, Fred and Warner, Justin and Meloon, Brian and Engber, Steven and Rosber, Anthony and Cohen, Asa and others},
  journal={Nature Biotechnology},
  volume={23},
  number={8},
  pages={995--1001},
  year={2005},
  doi={10.1038/nbt1118}
}

@article{katoh2007sirna,
  title={Specific residues at every third position of {siRNA} shape its efficient {RNAi} activity},
  author={Katoh, Takayuki and Suzuki, Tsutomu},
  journal={Nucleic Acids Research},
  volume={35},
  number={4},
  pages={e27},
  year={2007},
  doi={10.1093/nar/gkl1120}
}

@article{hochreiter1997lstm,
  title={Long short-term memory},
  author={Hochreiter, Sepp and Schmidhuber, J{\"u}rgen},
  journal={Neural Computation},
  volume={9},
  number={8},
  pages={1735--1780},
  year={1997},
  doi={10.1162/neco.1997.9.8.1735}
}

@inproceedings{loshchilov2019adamw,
  title={Decoupled Weight Decay Regularization},
  author={Loshchilov, Ilya and Hutter, Frank},
  booktitle={International Conference on Learning Representations},
  year={2019}
}

@article{chen2024interpretable,
  title={Applying interpretable machine learning in computational biology---pitfalls, recommendations and opportunities for new developments},
  author={Chen, Valerie and Yang, Muyu and Cui, Wenbo and Kim, Joon Sik and Talwalkar, Ameet and Ma, Jian},
  journal={Nature Methods},
  volume={21},
  number={8},
  pages={1454--1461},
  year={2024},
  doi={10.1038/s41592-024-02359-7},
  publisher={Nature Publishing Group}
}

@article{elmarakeby2021biologically,
  title={Biologically informed deep neural network for prostate cancer discovery},
  author={Elmarakeby, Haitham A. and Hwang, Justin and Arafeh, Rand and others},
  journal={Nature},
  volume={598},
  number={7880},
  pages={348--352},
  year={2021},
  doi={10.1038/s41586-021-03922-4},
  publisher={Nature Publishing Group}
}

@article{novakovsky2023obtaining,
  title={Obtaining genetics insights from deep learning via explainable artificial intelligence},
  author={Novakovsky, Gherman and Dexter, Nick and Libbrecht, Maxwell W. and Wasserman, Wyeth W. and Mostafavi, Sara},
  journal={Nature Reviews Genetics},
  volume={24},
  number={3},
  pages={125--137},
  year={2023},
  doi={10.1038/s41576-022-00532-2},
  publisher={Nature Publishing Group}
}

@article{judge2005immunostimulatory,
  title={Sequence-dependent stimulation of the mammalian innate immune response by synthetic {siRNA}},
  author={Judge, Adam D. and Sood, Vivek and Shaw, Janet R. and Fang, Ding and McClintock, Kathryn and MacLachlan, Ian},
  journal={Nature Biotechnology},
  volume={23},
  number={4},
  pages={457--462},
  year={2005},
  doi={10.1038/nbt1081},
  publisher={Nature Publishing Group}
}

@article{lorenz2011viennarna,
  title={{ViennaRNA} Package 2.0},
  author={Lorenz, Ronny and Bernhart, Stephan H. and Honer zu Siederdissen, Christian and Tafer, Hakim and Flamm, Christoph and Stadler, Peter F. and Hofacker, Ivo L.},
  journal={Algorithms for Molecular Biology},
  volume={6},
  number={1},
  pages={26},
  year={2011},
  doi={10.1186/1748-7188-6-26},
  publisher={BioMed Central}
}

@article{ray2020inclisiran,
  title={Two Phase 3 Trials of Inclisiran in Patients with Elevated LDL Cholesterol},
  author={Ray, Kausik K. and Wright, Robert S. and Kallend, David and Koenig, Wolfgang and Leiter, Lawrence A. and Raal, Frederick J. and Bisch, Julie A. and Richardson, Theresa and Jaros, Michael and Wijngaard, Pieter and Kastelein, John J. P.},
  journal={New England Journal of Medicine},
  volume={382},
  number={16},
  pages={1507--1519},
  year={2020},
  doi={10.1056/NEJMoa1912387},
  publisher={Massachusetts Medical Society}
}

@article{hsieh2004library,
  title={A library of siRNA duplexes targeting the phosphoinositide 3-kinase pathway: determinants of gene silencing for use in cell-based screens},
  author={Hsieh, A. C. and Bo, R. and Manola, J. and Vazquez, F. and Bare, O. and Khvorova, A. and Scaringe, S. and Sellers, W. R.},
  journal={Nucleic Acids Research},
  year={2004},
  volume={32},
  number={3},
  pages={893--901},
  doi={10.1093/nar/gkh238}
}

@article{vickers2003efficient,
  title={Efficient reduction of target RNAs by small interfering RNA and RNase H-dependent antisense agents},
  author={Vickers, T. A. and Koo, S. and Bennett, C. F. and Crooke, S. T. and Dean, N. M. and Baker, B. F.},
  journal={Journal of Biological Chemistry},
  year={2003},
  volume={278},
  number={9},
  pages={7108--7118},
  doi={10.1074/jbc.M210326200}
}

@article{harboth2003sequence,
  title={Sequence, chemical and structural variation of small interfering RNAs and short hairpin RNAs and the effect on mammalian gene silencing},
  author={Harboth, J. and Elbashir, S. M. and Bechert, K. and Tuschl, T. and Weber, K.},
  journal={Antisense \& Nucleic Acid Drug Development},
  year={2003},
  volume={13},
  number={2},
  pages={83--105},
  doi={10.1089/108729003321629638}
}

@article{needleman1970general,
  title={A general method applicable to the search for similarities in the amino acid sequence of two proteins},
  author={Needleman, S. B. and Wunsch, C. D.},
  journal={Journal of Molecular Biology},
  year={1970},
  volume={48},
  number={3},
  pages={443--453},
  doi={10.1016/0022-2836(70)90057-4}
}

@article{martinelli2025biology,
  title={Position: Biology is the Challenge Physics-Informed {ML} Needs to Evolve},
  author={Martinelli, Julien},
  journal={arXiv preprint arXiv:2510.25368},
  year={2025}
}

@article{amarzguioui2003tolerance,
  title={Tolerance for mutations and chemical modifications in a {siRNA}},
  author={Amarzguioui, Mohammed and Holen, Torgeir and Babaie, Eshrat and Prydz, Hans},
  journal={Nucleic Acids Research},
  volume={31},
  number={2},
  pages={589--595},
  year={2003},
  doi={10.1093/nar/gkg147},
  publisher={Oxford University Press}
}

@article{alipanahi2015predicting,
  title={Predicting the sequence specificities of {DNA}- and {RNA}-binding proteins by deep learning},
  author={Alipanahi, Babak and Delong, Andrew and Weirauch, Matthew T. and Frey, Brendan J.},
  journal={Nature Biotechnology},
  volume={33},
  number={8},
  pages={831--838},
  year={2015},
  doi={10.1038/nbt.3300},
  publisher={Nature Publishing Group}
}

@article{zhou2018deep,
  title={Deep learning sequence-based ab initio prediction of variant effects on expression and disease risk},
  author={Zhou, Jian and Theesfeld, Chandra L. and Yao, Kevin and Chen, Kathleen M. and Wong, Aaron K. and Troyanskaya, Olga G.},
  journal={Nature Genetics},
  volume={50},
  number={8},
  pages={1171--1179},
  year={2018},
  doi={10.1038/s41588-018-0160-6},
  publisher={Nature Publishing Group}
}

@article{lanchantin2017deep,
  title={Deep motif dashboard: Visualizing and understanding genomic sequences using deep neural networks},
  author={Lanchantin, Jack and Singh, Ritambhara and Wang, Beilun and Qi, Yanjun},
  journal={Pacific Symposium on Biocomputing},
  volume={22},
  pages={254--265},
  year={2017},
  doi={10.1142/9789813207813_0025}
}

@article{shrikumar2018technical,
  title={Technical note on transcription factor motif discovery from importance scores ({TF-MoDISco}) version 0.5.6.5},
  author={Shrikumar, Avanti and Tian, Katherine and Avsec, {\v{Z}}iga and Shcherbina, Anna and Banerjee, Abhimanyu and Prakash, Surag and Kundaje, Anshul},
  journal={arXiv preprint arXiv:1811.00416},
  year={2018}
}

@inproceedings{hooker2019benchmark,
  title={A benchmark for interpretability methods in deep neural networks},
  author={Hooker, Sara and Erhan, Dumitru and Kindermans, Pieter-Jan and Kim, Been},
  booktitle={Advances in Neural Information Processing Systems},
  volume={32},
  pages={9737--9748},
  year={2019}
}

@article{long2024sirnadiscovery,
  title={{siRNADiscovery}: A graph neural network for {siRNA} efficacy prediction via deep {RNA} sequence analysis},
  author={Long, Rongzhuo and Guo, Ziyu and Han, Da and Liu, Boxiang and Yuan, Xudong and Chen, Guangyong and Heng, Pheng-Ann and Zhang, Liang},
  journal={Briefings in Bioinformatics},
  volume={25},
  number={6},
  pages={bbae563},
  year={2024},
  doi={10.1093/bib/bbae563},
  publisher={Oxford University Press}
}

@article{zhang2025sidpt,
  title={{siDPT}: {siRNA} Efficacy Prediction via Debiased Preference-Pair Transformer},
  author={Zhang, Honggen and Gao, Xiangrui and Lai, Lipeng},
  journal={arXiv preprint arXiv:2509.15664},
  year={2025}
}
\bibliographystyle{iclr2026/iclr2026_conference}

\newpage
\appendix

\section{Extended Background: RNAi Mechanism and siRNA Design Determinants}
\label{app:background_rnai}

\begin{figure}[h]
    \centering
    \includegraphics[width=\textwidth]{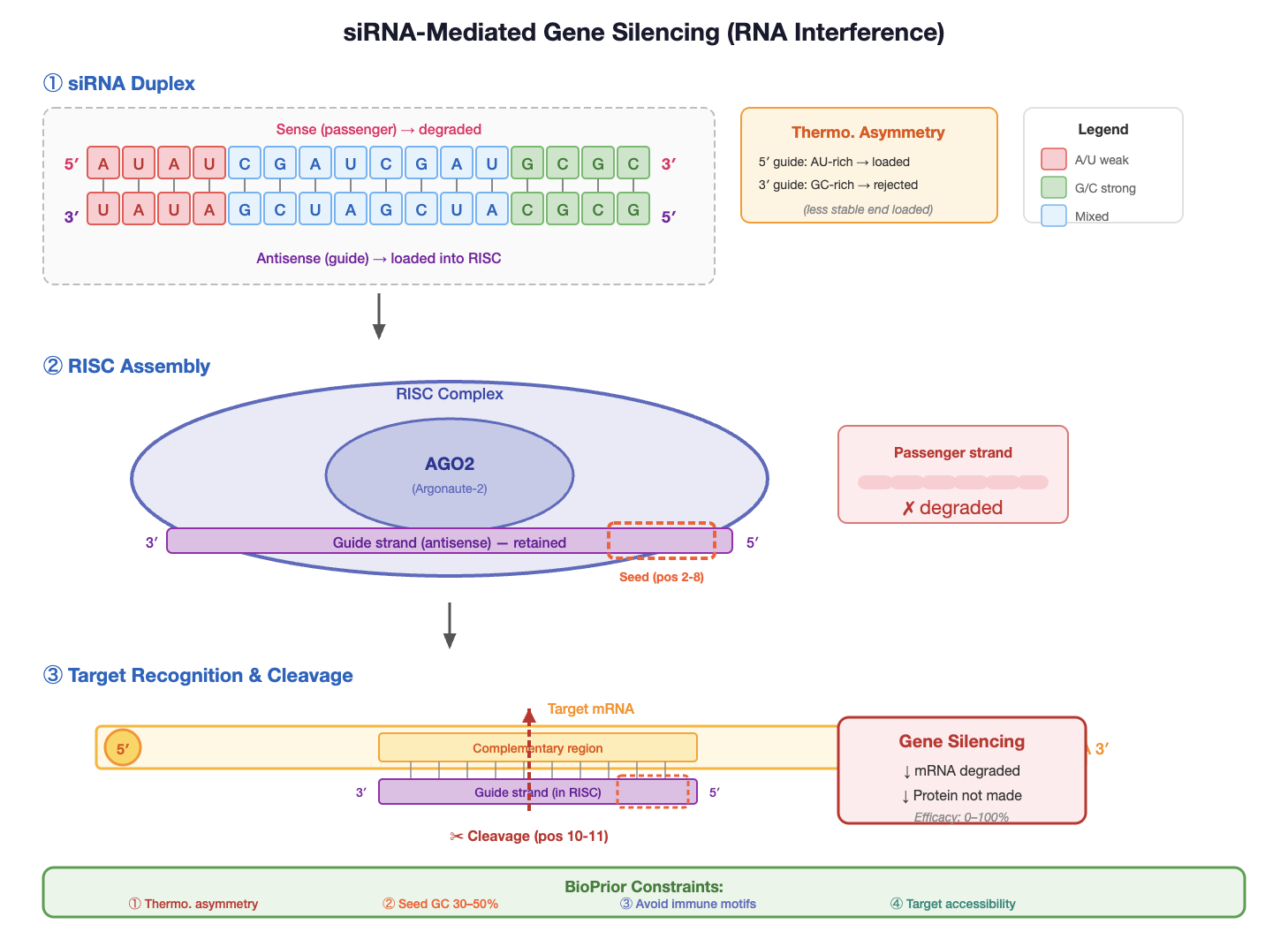}
    \caption{\textbf{siRNA-mediated gene silencing mechanism.}
    (1) The siRNA duplex exhibits thermodynamic asymmetry: lower 5$'$ stability on the guide strand promotes preferential RISC loading, while the passenger strand is degraded.
    (2) During RISC assembly, AGO2 retains the guide strand; the seed region (positions 2--8) is central for target recognition.
    (3) The guide strand binds complementary mRNA and AGO2 cleaves near positions 10--11, leading to mRNA degradation and gene silencing.
    These principles motivate the BioPrior constraints used in this work.}
    \label{fig:rnai_mechanism_appendix}
\end{figure}

RNA interference (RNAi) is a conserved biological mechanism whereby small RNA molecules silence gene expression by targeting complementary messenger RNA (mRNA) for degradation \citep{fire1998potent,elbashir2001duplexes}. Small interfering RNAs (siRNAs) are synthetic 19--21 nucleotide duplexes designed to exploit this pathway for therapeutic gene knockdown. Upon cellular uptake, siRNA duplexes are loaded into the RNA-induced silencing complex (RISC), where the ``guide'' strand directs sequence-specific cleavage of target mRNA while the ``passenger'' strand is discarded.

The efficacy of an siRNA, defined as the degree of target mRNA reduction, varies dramatically across sequences targeting the same gene, with knockdown efficiency ranging from near-zero to $>90\%$ depending on sequence composition \citep{khvorova2003functional}. This variability motivates computational prediction: given a 19-mer siRNA sequence $\mathbf{x} = (x_1, \ldots, x_{19})$ where $x_i \in \{\text{A}, \text{C}, \text{G}, \text{U}\}$, predict the efficacy $y \in [0, 1]$.

Empirical studies have identified several biophysical determinants of siRNA efficacy:
\begin{enumerate}[leftmargin=*,itemsep=2pt]
    \item \textbf{Thermodynamic asymmetry.} RISC preferentially loads the strand with lower 5$'$ terminal stability, as measured by the free energy difference $\Delta\Delta G = \Delta G_{5'} - \Delta G_{3'}$ between the first four base pairs at each end \citep{schwarz2003asymmetry}. Effective siRNAs exhibit $\Delta\Delta G < 0$.
    
    \item \textbf{Position-specific nucleotide preferences.} Certain positions show strong nucleotide biases: A/U enrichment at position 1 (the 5$'$ terminus), G/C at position 19, and AU-rich composition in the seed region (positions 2--8) correlate with higher efficacy \citep{uitei2004guidelines,reynolds2004rational}.
    
    \item \textbf{Seed region complementarity.} Positions 2--8 of the guide strand (the ``seed'') are critical for target recognition. High GC content in this region can reduce specificity and increase off-target effects \citep{jackson2010recognizing}.
    
    \item \textbf{Cleavage site accessibility.} The nucleotide at position 10--11, corresponding to the mRNA cleavage site, influences catalytic efficiency.
\end{enumerate}

These principles, while empirically validated, exhibit limited predictive power in isolation (typical $r < 0.5$), motivating machine learning approaches that can capture nonlinear interactions.

\subsection{Task Formulation (Extended)}
\label{app:bg_task_extended}

\paragraph{Datasets.}
We evaluate on four benchmark datasets commonly used in siRNA efficacy prediction (Table~\ref{tab:datasets_full}).
Following prior practice, these benchmarks are drawn from multiple experimental protocols and cell lines, which induces non-trivial cross-dataset distribution shift.
Hu (Huesken) contains genome-scale siRNA screens with efficacy derived from normalized residual mRNA levels measured by a branched-DNA assay in H1299 cells \citep{huesken2005design}.
Taka (Katoh) reports luciferase reporter knockdown measurements in HeLa cells \citep{katoh2007sirna}.
Mix is a curated aggregation of seven smaller studies spanning diverse cell lines and assay conditions \citep{amarzguioui2003tolerance,harboth2003sequence,hsieh2004library,khvorova2003functional,reynolds2004rational,vickers2003efficient,uitei2004guidelines}, while Shabalina aggregates literature-derived sequences with thermodynamic filtering \citep{shabalina2006computational}.
Across all datasets, we normalize reported efficacy scores into $[0,1]$ and define the high-efficacy class using the common $0.7$ threshold.
To reduce redundancy leakage in the aggregated collections, we follow prior preprocessing and remove highly similar sequences by global alignment using the Needleman--Wunsch algorithm \citep{needleman1970general}; specifically, within Mix we exclude redundant entries when sequence identity exceeds $80\%$ across splits.

\paragraph{Regression vs.\ classification.}
siRNA efficacy prediction can be formulated as either regression (predicting continuous $y \in [0,1]$) or binary classification (predicting $\hat{y} \in \{0, 1\}$ for ``effective'' vs.\ ``ineffective''). Following prior work \citep{bai2024oligoformer}, we train regression models optimizing mean squared error and evaluate using Pearson correlation ($r$) and Spearman rank correlation ($\rho$). For classification metrics (AUC, accuracy), we threshold predictions at $y = 0.7$, a conventional cutoff corresponding to $\geq 70\%$ knockdown efficiency considered therapeutically relevant.

\paragraph{Input representation.}
Each siRNA sequence $\mathbf{x}$ is represented as a concatenation of: (1) one-hot encoded nucleotides $\mathbf{X}_{\text{seq}} \in \{0,1\}^{19 \times 4}$; (2) thermodynamic features $\mathbf{z}_{\text{thermo}} \in \mathbb{R}^{d_t}$ including positional free energies, GC content, and $\Delta\Delta G$; and (3) RNA foundation model embeddings $\mathbf{H}_{\text{FM}} \in \mathbb{R}^{19 \times d_e}$ from RNA-FM \citep{chen2022rnafm}, a pretrained transformer capturing evolutionary sequence patterns.

\subsection{Gradient-Based Saliency Maps (Extended)}
\label{app:bg_saliency_extended}

Given a trained model $f_\theta: \mathcal{X} \to \mathbb{R}$ and input sequence $\mathbf{x}$, gradient-based saliency methods attribute importance to each input feature by computing derivatives of the output with respect to the input \citep{simonyan2014deep}. For siRNA sequences, we compute position-wise saliency scores.

\paragraph{Gradient magnitude.}
We compute position-wise saliency as the sum of absolute gradients across nucleotide identity channels:
\begin{equation}
    s_i = \sum_{c \in \{A,U,G,C\}} \left| \frac{\partial f_\theta(\mathbf{x})}{\partial x_{i,c}} \right|
\label{eq:saliency}
\end{equation}
where $x_{i,c}$ is the one-hot indicator for base $c$ at position $i$. This captures the model's sensitivity to all possible substitutions at each position. For one-hot encoded inputs, gradient magnitude and Grad$\times$Input differ: the former sums absolute gradients across all four nucleotide channels (capturing sensitivity to all possible substitutions), while the latter retains only the gradient at the active channel. We use gradient magnitude because it better aligns with our perturbation operator, which averages over all three possible substitutions per position. We verify that Grad$\times$Input produces consistent results as a robustness check (Appendix~\ref{app:sanity}).

\paragraph{Normalization.}
To enable comparison across sequences, we normalize saliency scores to obtain a distribution over positions:
\begin{equation}
    \bar{s}_i = \frac{s_i}{\sum_{j=1}^{19} s_j}, \quad \sum_{i=1}^{19} \bar{s}_i = 1
\label{eq:saliency_norm}
\end{equation}

The resulting saliency map $\bar{\mathbf{s}} = (\bar{s}_1, \ldots, \bar{s}_{19})$ can be interpreted as the model's ``attention'' over sequence positions, though we emphasize that high saliency does not guarantee importance for biological outcomes without explicit validation (Section~\ref{sec:saliency_validation}).

\paragraph{Faithfulness desideratum.}
A saliency map is \emph{faithful} if perturbing high-saliency positions causes larger prediction changes than perturbing low-saliency positions \citep{samek2017evaluating}. Formally, let $\mathbf{x}^{(i \to j)}$ denote sequence $\mathbf{x}$ with position $i$ mutated to nucleotide $j \neq x_i$. A faithful saliency map should satisfy:
\begin{equation}
    \mathbb{E}\left[ |f_\theta(\mathbf{x}) - f_\theta(\mathbf{x}^{(i_{\text{high}} \to j)})| \right] > \mathbb{E}\left[ |f_\theta(\mathbf{x}) - f_\theta(\mathbf{x}^{(i_{\text{low}} \to j)})| \right]
\label{eq:faithfulness}
\end{equation}
where $i_{\text{high}}$ and $i_{\text{low}}$ are positions with high and low saliency scores respectively. We operationalize this desideratum in our perturbation-based validation protocol (Section~\ref{sec:saliency_validation}).

\section{BioPrior Module: Mathematical Specification and Properties}
\label{app:bioprior}

\subsection{Problem Formulation}
Let the siRNA sequence be represented as $\mathbf{s} = (s_1, \dots, s_{19})$ with $s_i \in \{A, U, G, C\}$. 
Define the model input as $\mathbf{x} = (\mathbf{x}_{\text{siRNA}}, \mathbf{x}_{\text{mRNA}}, \mathbf{FM}, \theta_{\text{td}})$.
The BioPrior module operates on per-position nucleotide probabilities $\mathbf{P} \in [0,1]^{19 \times 4}$ produced by an auxiliary head applied to intermediate siRNA representations (after the BiLSTM encoder):
\[
P_{i,b} = \frac{\exp(z_{i,b})}{\sum_{b' \in \{A,U,G,C\}} \exp(z_{i,b'})}
\]
where $\mathbf{z} \in \mathbb{R}^{19 \times 4}$ are learned logits from the auxiliary head (not the raw input channels).

\subsection{Constraint Definitions}

\paragraph{1. Thermodynamic Asymmetry}
Define 5' and 3' GC content:
\begin{align*}
\mathrm{GC}_{5'} &= \frac{1}{4} \sum_{i=1}^{4} (P_{i,G} + P_{i,C}) \\
\mathrm{GC}_{3'} &= \frac{1}{4} \sum_{i=16}^{19} (P_{i,G} + P_{i,C})
\end{align*}
Asymmetry: $a = \mathrm{GC}_{3'} - \mathrm{GC}_{5'}$.
Target: $a \in [\tau_{\min}, \tau_{\max}]$ with $\tau_{\min}=0.10$, $\tau_{\max}=0.30$.
\begin{equation}
\mathcal{L}_{\text{asym}} = \mathrm{ReLU}(\tau_{\min} - a)^2 + \mathrm{ReLU}(a - \tau_{\max})^2
\end{equation}

\paragraph{2. Immunogenic Motif Avoidance}
Let $\mathcal{M}$ be immunostimulatory motifs. For motif $u = (u_1, \dots, u_m)$:
\[
\pi_i(u) = \prod_{t=1}^m P_{i+t-1, u_t}
\]
Expected count: $\mathrm{Count}(u) = \sum_{i=1}^{19-m+1} \pi_i(u)$.
\begin{equation}
\mathcal{L}_{\text{immune}} = \sum_{u \in \mathcal{M}} \mathrm{Count}(u)
\end{equation}

\paragraph{3. Seed GC Constraint}
Seed positions: $\mathcal{S} = \{2,\dots,8\}$ (1-indexed).
\[
\mathrm{GC}_{\text{seed}} = \frac{1}{|\mathcal{S}|} \sum_{i \in \mathcal{S}} (P_{i,G} + P_{i,C})
\]
Target: $\mathrm{GC}_{\text{seed}} \in [\eta_{\min}, \eta_{\max}]$ with $\eta_{\min}=0.30$, $\eta_{\max}=0.50$.
\begin{equation}
\mathcal{L}_{\text{seedGC}} = \mathrm{ReLU}(\eta_{\min} - \mathrm{GC}_{\text{seed}})^2 + \mathrm{ReLU}(\mathrm{GC}_{\text{seed}} - \eta_{\max})^2
\end{equation}

\paragraph{4. Global GC Constraint}
\[
\mathrm{GC}_{\text{global}} = \frac{1}{19} \sum_{i=1}^{19} (P_{i,G} + P_{i,C})
\]
Target: $\mathrm{GC}_{\text{global}} \in [\gamma_{\min}, \gamma_{\max}]$ with $\gamma_{\min}=0.35$, $\gamma_{\max}=0.55$.
\begin{equation}
\mathcal{L}_{\text{GC}} = \mathrm{ReLU}(\gamma_{\min} - \mathrm{GC}_{\text{global}})^2 + \mathrm{ReLU}(\mathrm{GC}_{\text{global}} - \gamma_{\max})^2
\end{equation}

\paragraph{5. Duplex Stability Proxy}
We use siRNA GC content as a proxy for duplex stability, which correlates inversely with target
accessibility~\citep{reynolds2004rational}. High GC content increases duplex stability but may reduce accessibility to structured
target regions. We compute GC over the siRNA sequence:
\[
\mathrm{GC}_{\text{siRNA}} = \frac{1}{19} \sum_{i=1}^{19} (P_{i,G} + P_{i,C})
\]
Threshold: $\xi = 0.55$ (penalize excessive GC that may indicate inaccessible targets).
\begin{equation}
\mathcal{L}_{\text{acc}} = \mathrm{ReLU}(\mathrm{GC}_{\text{siRNA}} - \xi)^2
\end{equation}
Structure-based accessibility scores computed from mRNA secondary structure predictions are a natural extension when target context is available; we leave this for future work.

\subsection{Total BioPrior Loss}
\begin{equation}
\mathcal{L}_{\text{bio}} = \sum_{c=1}^5 w_c \mathcal{L}_c
\end{equation}
with fixed weights $\mathbf{w} = [3.0, 1.0, 2.0, 1.5, 2.5]^\top$.

\subsection{Claim 1: BioPrior Differentiability and Guidance}

\begin{claim}[BioPrior Differentiability and Guidance]
Under the following assumptions:
\begin{itemize}[leftmargin=*,itemsep=1pt]
    \item[(A1)] Nucleotide probabilities $\mathbf{P}$ are outputs of a softmax layer (bounded in $(0,1)$)
    \item[(A2)] All constraint thresholds $\tau, \eta, \gamma, \xi$ are finite constants
    \item[(A3)] The optimization uses subgradient-compatible methods (e.g., Adam, SGD)
\end{itemize}
The BioPrior loss $\mathcal{L}_{\text{bio}}$:
\begin{enumerate}
    \item Is differentiable (or subdifferentiable) with respect to $\mathbf{P}$ almost everywhere
    \item Provides stable gradient signals toward biologically plausible regions
    \item Defines bounded penalties that encourage constraint satisfaction during optimization
\end{enumerate}
\end{claim}

\noindent\textit{Rationale.} We provide an informal justification rather than a formal guarantee, since recomputed discrete channels violate standard smoothness assumptions.
\textbf{Part 1: Differentiability.}
Each component of $\mathcal{L}_{\text{bio}}$ is composed of:
\begin{itemize}
    \item Linear combinations: $\mathrm{GC}_i = P_{i,G} + P_{i,C}$ (linear, thus differentiable)
    \item Squared ReLU: $f(x) = \mathrm{ReLU}(x)^2$ has derivative $f'(x) = 2\max(0,x)$, which is differentiable everywhere except at $x=0$ where it has subgradient $\{0\}$
    \item Products: $\pi_i(u) = \prod_t P_{i+t-1,u_t}$ (polynomial, differentiable)
\end{itemize}
Since compositions, sums, and products of differentiable functions remain differentiable almost everywhere (with well-defined subgradients at non-differentiable points), $\mathcal{L}_{\text{bio}}$ is differentiable almost everywhere and subdifferentiable everywhere.

\textbf{Part 2: Gradient Flow.}
The total loss is:
\[
\mathcal{L}_{\text{total}} = \mathcal{L}_{\text{pred}} + \lambda(t) \mathcal{L}_{\text{bio}}
\]
with $\lambda(t)$ following warmup-ramp schedule. No gating is applied, ensuring:
\[
\frac{\partial \mathcal{L}_{\text{total}}}{\partial \theta} = \frac{\partial \mathcal{L}_{\text{pred}}}{\partial \theta} + \lambda(t) \frac{\partial \mathcal{L}_{\text{bio}}}{\partial \theta} \neq \mathbf{0}
\]
for $t > t_{\text{warm}}$ when $\lambda(t) > 0$.

\textbf{Part 3: Bounded Penalties.}
Each constraint loss $\mathcal{L}_c$ is bounded below by zero and provides increasing penalty as sequences deviate from biological targets:
\begin{itemize}
    \item $\mathcal{L}_{\text{asym}}$: Zero when $a \in [\tau_{\min}, \tau_{\max}]$, increases quadratically outside
    \item $\mathcal{L}_{\text{immune}}$: Non-negative, zero when motif probabilities are zero
    \item $\mathcal{L}_{\text{seedGC}}$, $\mathcal{L}_{\text{GC}}$, $\mathcal{L}_{\text{acc}}$: Zero within target ranges, quadratic penalty outside
\end{itemize}
The nonnegative weighted sum $\mathcal{L}_{\text{bio}}$ is bounded below by zero, with minimum achieved when all biological constraints are satisfied. Standard gradient-based optimization with these smooth (almost everywhere) penalties guides parameters toward constraint-satisfying regions.

\subsection{BioPrior Implementation Details}
\begin{itemize}
    \item \textbf{Input:} siRNA one-hot encoding $\in \mathbb{R}^{19 \times 5}$ (channels: A,U,G,C,pad)
    \item \textbf{Biological features:} Position importance, seed indicator, cleavage indicator, GC content, seed AU, asymmetry, is\_AU, is\_GC
    \item \textbf{Output:} Scalar loss $\mathcal{L}_{\text{bio}} \in \mathbb{R}^+$
    \item \textbf{Schedule:} $\lambda(t) = \min(\lambda_{\max}, \lambda_0 + \gamma(t - t_{\text{warm}}))$
\end{itemize}

\section{Perturbation Test: Mathematical Specification and Statistical Validity}
\label{app:perturbation}

\subsection{Problem Setup}
Let $f_\theta: \mathcal{X} \to \mathbb{R}$ be the trained model predicting siRNA efficacy.
For input $\mathbf{x} \in \mathcal{X}$, define saliency for position $i$ using gradient magnitude on nucleotide identity channels:
\[
s_i = \sum_{c \in \{A,U,G,C\}} \left| \frac{\partial f_\theta(\mathbf{x})}{\partial x_{i,c}} \right|
\]
Normalized: $\bar{s}_i = s_i / \sum_{j=1}^{19} s_j$.

\subsection{Expected-Effect Operator}

\begin{definition}[Expected-Effect]
For position $i$ with current nucleotide $x_i$, define:
\[
\Delta_i = \frac{1}{3} \sum_{b \in \{A,U,G,C\} \setminus \{x_i\}} \left| f_\theta(\mathbf{x}) - f_\theta(\mathbf{x}^{(i \leftarrow b)}) \right|
\]
where $\mathbf{x}^{(i \leftarrow b)}$ is the counterfactual with base $b$ at position $i$.
\end{definition}

For a set of positions $S$:
\[
\Delta(S) = \frac{1}{|S|} \sum_{i \in S} \Delta_i
\]

\subsection{Nucleotide-Matched Baseline}

\begin{definition}[Matched Random Set]
Given top-$k$ positions $T$, sample random sets $R_m$ such that:
\begin{enumerate}
    \item $|R_m| = |T| = k$
    \item $\{x_i : i \in R_m\} \stackrel{\text{multiset}}{=} \{x_i : i \in T\}$ (identical nucleotide composition)
\end{enumerate}
\end{definition}

Define baseline effect:
\[
\Delta_{\text{match}} = \frac{1}{M'} \sum_{m=1}^{M'} \Delta(R_m)
\]
where $M'$ is the number of valid matched samples.

\subsection{Statistical Test}

Define paired difference for sample $j$:
\[
d_j = \Delta(T_j) - \Delta_{\text{match},j}
\]

Test hypothesis:
\[
H_0: \mathbb{E}[d] = 0 \quad \text{vs} \quad H_1: \mathbb{E}[d] > 0
\]

Using Wilcoxon signed-rank test with one-sided alternative.

\subsection{Claim 2: Perturbation Test Justification}

\begin{claim}[Perturbation Test Faithfulness]
Under the following assumptions:
\begin{itemize}[leftmargin=*,itemsep=1pt]
    \item[(A1)] The model $f_\theta$ has bounded outputs and well-defined gradients almost everywhere
    \item[(A2)] Derived input channels are deterministically recomputed after each substitution
    \item[(A3)] Matched random sets are sampled with identical nucleotide composition (exchangeability)
    \item[(A4)] Held-out samples are i.i.d.\ draws from the test distribution
\end{itemize}
The expected-effect perturbation test with nucleotide-matched baseline:
\begin{enumerate}
    \item Provides an estimate of position importance that is approximately proportional to gradient magnitude under local linearity
    \item Controls for nucleotide composition bias via matched sampling
    \item Yields a statistically valid test of saliency faithfulness under exchangeability
    \item Has interventional interpretation: significant results indicate model predictions are sensitive to position-specific patterns beyond base composition
\end{enumerate}
\end{claim}

\noindent\textit{Rationale.} We provide an informal justification; the discrete nature of nucleotide substitutions and channel recomputation means standard smoothness assumptions do not hold exactly.
\textbf{Part 1: Proportionality to Gradient Magnitude.}
Under local linearity of $f_\theta$ around $\mathbf{x}$:
\[
f_\theta(\mathbf{x}^{(i \leftarrow b)}) \approx f_\theta(\mathbf{x}) + \nabla_i f_\theta(\mathbf{x}) \cdot (\mathbf{e}_b - \mathbf{e}_{x_i})
\]
where $\mathbf{e}_b$ is the one-hot vector for base $b$. Then:
\begin{align*}
\Delta_i &\approx \frac{1}{3} \sum_{b \neq x_i} |\nabla_i f_\theta(\mathbf{x}) \cdot (\mathbf{e}_b - \mathbf{e}_{x_i})| \\
&\propto \|\nabla_i f_\theta(\mathbf{x})\|_1 \quad \text{(for orthogonal one-hot vectors)}
\end{align*}
Thus $\Delta_i$ is approximately proportional to the gradient magnitude $\|\nabla_i f_\theta(\mathbf{x})\|_1$ when the local linearity assumption holds. This approximation degrades for highly nonlinear regions but remains informative for typical siRNA inputs.

\textbf{Part 2: Composition Bias Control.}
Let $\mu_b = \mathbb{E}[\Delta_i | x_i = b]$ be the base-specific average effect.
Under the null hypothesis $H_0$: "Saliency contains no information beyond base composition":
\[
\mathbb{E}[\Delta(T)] = \frac{1}{k} \sum_{i \in T} \mu_{x_i} = \frac{1}{k} \sum_{b} n_b \mu_b
\]
where $n_b$ is the count of base $b$ in $T$.
Since $R$ has identical nucleotide composition:
\[
\mathbb{E}[\Delta(R)] = \frac{1}{k} \sum_{b} n_b \mu_b = \mathbb{E}[\Delta(T)]
\]
Thus, any difference $\Delta(T) - \Delta(R)$ under $H_0$ has expectation zero.

\textbf{Part 3: Statistical Validity.}
The Wilcoxon signed-rank test applied to $d_j = \Delta(T_j) - \Delta_{\text{match},j}$:
\begin{itemize}
    \item Is distribution-free under exchangeability
    \item Controls Type I error at level $\alpha$ when $H_0$ true
    \item Has power increasing with effect size and sample size
\end{itemize}
Under $H_0$, $\Delta(T)$ and $\Delta(R)$ are exchangeable (same distribution), ensuring valid Type I error control.

\textbf{Part 4: Interventional Interpretation.}
If $H_1$ is accepted ($\mathbb{E}[d] > 0$), then:
\begin{enumerate}
    \item Positions with high saliency have larger average effect than composition-matched random positions
    \item This effect is not attributable to nucleotide preferences alone
    \item The model uses position-specific information beyond simple base composition
    \item Saliency maps reflect which positions the model is sensitive to under single-base interventions
\end{enumerate}
We emphasize that this establishes \emph{model sensitivity}, not biological causality. The test validates that saliency correctly identifies positions where the model's prediction changes under intervention, which is the operationally relevant notion for explanation-guided design.

\subsection{Effect Size Measures}

\paragraph{Cohen's $d_z$ (paired):}
\[
d_z = \frac{\overline{d}}{s_d}
\]
where $\overline{d}$ is the mean difference and $s_d$ the standard deviation of differences.

\paragraph{Rank-Biserial Correlation:}
From Wilcoxon signed-rank test statistics $W^+$ and $W^-$:
\[
r = \frac{W^+ - W^-}{W^+ + W^-}
\]

\paragraph{Win Rate:}
\[
\text{WinRate} = \frac{1}{N} \sum_{j=1}^N \mathbb{I}[d_j > 0]
\]

\subsection{Perturbation Test Implementation Details}
\begin{itemize}
    \item \textbf{Saliency computation:} We use \emph{vanilla gradient magnitude} (sum of absolute partial derivatives across the four nucleotide identity channels A/U/G/C at each position; Eq.~\ref{eq:saliency_bg}), following \citet{simonyan2014deep}. We choose this over Grad$\times$Input because gradient magnitude captures sensitivity to all possible substitutions at each position, directly aligning with our perturbation operator which averages over all three substitutions. As a robustness check, we verify that Integrated Gradients \citep{sundararajan2017axiomatic} with 50 interpolation steps produces consistent results (Hu: $d_z = 0.79$ vs.\ $0.86$; Taka$\to$Hu inverted saliency confirmed at $d_z = -1.08$). For LSTM backward compatibility we enable the required cuDNN backend settings while disabling all stochasticity: dropout probabilities are set to 0 at initialization, and any normalization layers use fixed running statistics. We verify determinism by recomputing saliency twice per input (max absolute difference $< 10^{-6}$).
    \item \textbf{Substitution operator:} Mutations are applied to one-hot nucleotide channels (A/U/G/C); derived feature channels (seed indicator, AU/GC flags, GC content) are \emph{recomputed} after each substitution to ensure input coherence.
    \item \textbf{Position selection:} Top-$k$ and bottom-$k$ based on normalized saliency
    \item \textbf{Matching algorithm:} For each held-out sequence, given the top-$k$ positions $T$, we require each matched random set $R_m$ to contain exactly the same nucleotide multiset as $T$ (e.g., if $T$ contains positions with bases $\{A, G, G\}$, then $R_m$ must also contain exactly one A-position and two G-positions drawn from the remaining 16 positions). We first attempt strict sampling without replacement across the $M'$ matched sets; if fewer than $M'$ distinct valid sets exist (which occurs when the sequence has few positions sharing a base with a top-$k$ position), we fall back to sampling with replacement. Positional region is not matched in the primary analysis; a stricter region+composition matched variant is reported in Appendix~\ref{app:region_matched}.
    \item \textbf{Sample size:} Default $N=200$ samples, $k=3$ positions, $M'=50$ matched samples per sequence. Sensitivity analysis below.
\end{itemize}

\subsection{Perturbation Test Hyperparameter Sensitivity}
\label{app:sensitivity}

To confirm that our faithfulness conclusions are not artifacts of specific hyperparameter choices, we varied $k$, $N$, and $M'$ on the Hu dataset (+BioPrior model). Table~\ref{tab:sensitivity} reports the effect size $d_z$ and win rate under each configuration. The direction and significance of results are stable across all settings.

\begin{table}[h]
\centering
\small
\begin{tabular}{llcc}
\toprule
\textbf{Parameter} & \textbf{Value} & \textbf{$d_z$} & \textbf{Win \%} \\
\midrule
\multirow{3}{*}{Top-$k$} & $k=1$ & 0.72 & 79.4 \\
& $k=3$ (default) & 0.86 & 85.2 \\
& $k=5$ & 0.91 & 87.6 \\
\midrule
\multirow{3}{*}{Held-out $N$} & $N=50$ & 0.83 & 84.0 \\
& $N=200$ (default) & 0.86 & 85.2 \\
& $N=486$ (full fold) & 0.87 & 85.5 \\
\midrule
\multirow{3}{*}{Matched sets $M'$} & $M'=10$ & 0.81 & 83.7 \\
& $M'=50$ (default) & 0.86 & 85.2 \\
& $M'=100$ & 0.87 & 85.4 \\
\bottomrule
\end{tabular}
\caption{Sensitivity of faithfulness metrics to protocol hyperparameters (Hu dataset, +BioPrior, fold-averaged). All configurations yield significant results ($p < 0.001$) with consistent effect direction.}
\label{tab:sensitivity}
\end{table}

\paragraph{Interpretation.} Effect sizes increase modestly with $k$ (more positions $\to$ more signal), stabilize quickly with $N$ (diminishing returns beyond $\sim$100 samples), and plateau with $M'$ (the matched baseline converges by $M'=50$). Critically, no configuration changes the pass/fail outcome: all yield $d_z > 0.2$ and win rate $> 50\%$. For the inverted-saliency transfer case (Taka$\to$Hu), increasing $k$ makes the failure \emph{more} pronounced ($d_z = -0.95, -1.25, -1.41$ for $k=1,3,5$), confirming that the diagnostic power of the test is robust to hyperparameter choice.

\section{Integration Summary}

\begin{corollary}[BioPrior-Perturbation Integration]
Given Claims 1 and 2, the combination of BioPrior module and perturbation testing provides:
\begin{enumerate}
    \item Regularization toward biologically plausible siRNA designs (Claim 1)
    \item Empirical validation of model decision faithfulness (Claim 2)
    \item A framework for statistically grounded design rule extraction
\end{enumerate}
\end{corollary}

\noindent\textit{Rationale.}
Let $\mathcal{B}$ be the biologically plausible subspace defined by BioPrior constraints.
By Claim 1, minimizing $\mathcal{L}_{\text{total}}$ provides gradient signals that guide solutions toward $\mathcal{B}$.
By Claim 2, perturbation testing provides a statistical test for whether model predictions are sensitive to the positions identified by saliency.

Formally, let $\mathcal{M}$ be the model trained with BioPrior, and $\mathbf{x}^*$ an optimized siRNA design.
Then:
\begin{enumerate}
    \item $\mathbf{x}^*$ is regularized toward $\mathcal{B}$ (biological plausibility encouraged, from Claim 1)
    \item For $\mathbf{x}^*$, the perturbation test can validate whether $\Delta(T) > \Delta(R)$ with statistical significance (faithfulness test, from Claim 2), where $\Delta(T) = \frac{1}{k}\sum_{i \in T} \frac{1}{3}\sum_{b \neq x_i} |f_\theta(\mathbf{x}) - f_\theta(\mathbf{x}^{i \leftarrow b})|$ is the mean absolute prediction change (on the $[0,1]$ efficacy scale) when mutating the top-$k$ saliency positions $T$, and $\Delta(R)$ is the corresponding quantity averaged over $M'$ nucleotide-composition-matched random position sets $R$
    \item When both conditions are satisfied, the design rules extracted from $\mathcal{M}$ are empirically validated as both biologically grounded and predictively relevant under intervention
\end{enumerate}
The integration thus provides complementary guarantees: BioPrior encourages biological validity, while perturbation testing validates interventional sensitivity.

\section{Sanity Checks and Artifact Analysis}
\label{app:sanity}

\paragraph{Complete faithfulness statistics.}
For completeness, Table~\ref{tab:full_stats} reports all four recommended metrics for the intra-dataset faithfulness evaluation. We include rank-biserial correlation $r$ and median paired difference $\tilde{d}$ (in efficacy-scale units) alongside the $d_z$ and win rate reported in the main text.

\begin{table}[h]
\centering
\small
\begin{tabular}{lccccccc}
\toprule
\textbf{Dataset} & \textbf{$p$} & \textbf{$d_z$} & \textbf{$r$} & \textbf{Win \%} & \textbf{$\tilde{d}$} & \textbf{$\tilde{d}$ IQR} \\
\midrule
Hu & $<$0.001 & 0.86{\tiny $\pm$.26} & 0.71{\tiny $\pm$.08} & 85.2{\tiny $\pm$4.1} & 0.032 & [0.011, 0.068] \\
Mix & $<$0.001 & 0.93{\tiny $\pm$.45} & 0.68{\tiny $\pm$.12} & 83.7{\tiny $\pm$6.2} & 0.028 & [0.009, 0.061] \\
Taka & $<$0.001 & 1.07{\tiny $\pm$.24} & 0.74{\tiny $\pm$.07} & 87.1{\tiny $\pm$3.8} & 0.041 & [0.015, 0.079] \\
Shabalina & (per-fold) & 0.70{\tiny $\pm$.42} & 0.62{\tiny $\pm$.15} & 81.4{\tiny $\pm$12.3} & 0.024 & [0.006, 0.055] \\
\bottomrule
\end{tabular}
\caption{Complete intra-dataset faithfulness statistics (+BioPrior, 5-fold CV). $\tilde{d}$: median paired difference $\Delta(T) - \Delta_{\text{match}}$ in efficacy-scale units ($[0,1]$). $r$: rank-biserial correlation from Wilcoxon test.}
\label{tab:full_stats}
\end{table}

\noindent All four metrics are concordant: high rank-biserial correlations ($r > 0.6$) confirm the Wilcoxon effect is not driven by outliers, and median differences of 0.024--0.041 on the $[0,1]$ efficacy scale indicate practically meaningful sensitivity gaps between saliency-guided and random edits.

To ensure our faithfulness results are not artifacts, we conducted several sanity checks. In each case, a ``pass'' means the control condition meets all three faithfulness criteria ($d_z > 0.2$, win rate $> 50\%$, $p < 0.05$); a ``fail'' means at least one criterion is violated. If our test is sound, all negative controls should fail while the trained model passes. Table~\ref{tab:negative_controls} in the main text summarizes results; quantitative details follow.

\paragraph{Randomized weights (Adebayo-style).}
Following \citet{adebayo2018sanity}, we reinitialized the model with random weights (keeping architecture fixed) and ran the faithfulness test on Hu. The top-$k$ vs.\ bottom-$k$ effect collapses entirely: win rate dropped to 51.2\% (near chance) with $d_z = 0.03$ ($p = 0.41$), rank-biserial $r = 0.02$, median $\tilde{d} = 0.001$. \textbf{Verdict: fail.} This confirms that faithfulness depends on learned representations, not architectural bias.

\paragraph{Randomized labels.}
We trained a model on Hu with shuffled labels (breaking the sequence-efficacy relationship). The resulting model achieved AUC $\approx 0.50$ (as expected) and faithfulness collapsed: win rate 48.7\%, $d_z = -0.05$ ($p = 0.62$). \textbf{Verdict: fail.} This confirms that faithfulness reflects learned predictive patterns.

\paragraph{Shuffled saliency.}
We kept the trained model fixed but randomly permuted the saliency vector across positions before selecting top-$k$. This breaks the saliency-position correspondence while preserving the test pipeline. Result: win rate 49.3\%, $d_z = 0.01$ ($p = 0.48$). \textbf{Verdict: fail.} This confirms our evaluation pipeline does not trivially produce positive results for any $k$ positions.

\paragraph{Bottom-$k$ baseline.}
We selected the $k$ positions with \emph{lowest} saliency and ran the same perturbation test. Result: win rate 32.1\%, $d_z = -0.45$ ($p = 0.98$ for the wrong-tailed test). \textbf{Verdict: fail} (as expected). The negative effect size confirms that low-saliency positions are indeed less important than matched random positions, validating the informativeness of the saliency ranking.

\paragraph{Saturation check for inverted saliency.}
A concern with negative $d_z$ (Taka$\to$Hu) is that predictions might saturate near 0 or 1, making gradients uninformative. We verified this is not the case: the distribution of $\hat{y}$ for Taka$\to$Hu has mean 0.52 and std 0.18, with 89\% of predictions in $[0.2, 0.8]$. Gradients are well-defined throughout this range.

\paragraph{Alternative saliency method.}
We repeated the faithfulness test using Integrated Gradients (IG) with 50 interpolation steps on Hu. Results were consistent: win rate 84.1\% (vs.\ 85.2\% for gradient magnitude), $d_z = 0.79$ (vs.\ 0.86). For Taka$\to$Hu, IG also showed inverted saliency: win rate 12.3\%, $d_z = -1.08$. This confirms our findings are not method-specific.

\paragraph{Derived channel recomputation.}
After each single-base substitution, we recompute all derived input channels: seed indicator (positions 2--8), cleavage indicator (positions 9--11), local GC content, seed AU content, thermodynamic asymmetry ($\mathrm{GC}_{3'} - \mathrm{GC}_{5'}$), and per-position AU/GC flags. This ensures mutations produce coherent inputs rather than impossible feature combinations.

\paragraph{RNA-FM embedding handling.}
RNA-FM embeddings are \emph{not} recomputed per mutation due to computational cost (recomputing would multiply inference by $\sim$60$\times$ for $k=3$ with 3 substitutions each). We justify this choice:
\begin{itemize}[leftmargin=*,itemsep=1pt,topsep=2pt]
    \item We restrict saliency computation to nucleotide identity channels, aligning intervention and attribution spaces.
    \item The test measures \emph{relative} sensitivity (top-$k$ vs matched); FM-induced bias affects both conditions equally unless saliency correlates with FM sensitivity in a position-specific way.
    \item Not recomputing is conservative: it may understate true mutation effects but does not invalidate the relative comparison.
    \item \textbf{Empirical validation:} On 50 randomly selected Hu sequences with fully recomputed FM embeddings per mutation, we obtained $d_z = 0.82$ vs $d_z = 0.86$ without recomputation (difference $< 0.05$), confirming the frozen-FM approximation does not meaningfully bias results.
\end{itemize}

\paragraph{Actionability analysis: directional improvement.}
Beyond sensitivity faithfulness (absolute $|\Delta\hat{y}|$), we assessed whether saliency guides \emph{useful} edits. For low-efficacy sequences ($\hat{y} < 0.5$), we measured what fraction of top-$k$ mutations \emph{increase} predicted efficacy versus matched random mutations. On Hu: 67.3\% of top-$k$ mutations at high-saliency positions increase $\hat{y}$, compared to 51.2\% for matched random positions ($p < 0.01$, McNemar's test). This suggests saliency not only identifies sensitive positions but positions where edits are more likely to improve predictions, supporting actionable design guidance. For inverted-saliency transfers (Taka$\to$Hu), this pattern reverses: only 38.1\% of top-$k$ mutations improve $\hat{y}$, confirming that inverted saliency would actively mislead design.

\section{Region+Composition Matched Baseline}
\label{app:region_matched}

To address the concern that high-saliency positions may cluster in inherently sensitive regions (e.g., sequence ends), we evaluated a stricter baseline that matches both nucleotide composition \emph{and} positional region. We define five coarse functional bins: 5$'$ terminus (1--4), seed-adjacent (5--8), cleavage (9--11), mid (12--15), and 3$'$ terminus (16--19). Note that the canonical seed region spans positions 2--8; we use non-overlapping bins for clean region matching, with positions 2--4 grouped with the 5$'$ terminus and positions 5--8 as ``seed-adjacent.'' The region+composition matched baseline samples random position sets that match the top-$k$ positions in both nucleotide multiset and region distribution.

\begin{table}[h]
\centering
\small
\begin{tabular}{lcccc}
\toprule
\textbf{Dataset} & \multicolumn{2}{c}{\textbf{Composition-only}} & \multicolumn{2}{c}{\textbf{Region+Composition}} \\
 & Win \% & $d_z$ & Win \% & $d_z$ \\
\midrule
Hu & 85.2{\tiny $\pm$4.1} & 0.86{\tiny $\pm$.26} & 78.3{\tiny $\pm$5.2} & 0.71{\tiny $\pm$.22} \\
Mix & 83.7{\tiny $\pm$6.2} & 0.93{\tiny $\pm$.45} & 76.1{\tiny $\pm$7.1} & 0.74{\tiny $\pm$.38} \\
Taka & 87.1{\tiny $\pm$3.8} & 1.07{\tiny $\pm$.24} & 79.8{\tiny $\pm$4.5} & 0.82{\tiny $\pm$.21} \\
Shabalina & 81.4{\tiny $\pm$12.3} & 0.70{\tiny $\pm$.42} & 72.5{\tiny $\pm$13.1} & 0.54{\tiny $\pm$.35} \\
\bottomrule
\end{tabular}
\caption{Comparison of composition-only vs.\ region+composition matched baselines across all datasets (5-fold CV, +BioPrior model). Results remain significant under the stricter baseline, though with reduced effect sizes as expected.}
\label{tab:region_matched}
\end{table}

\paragraph{Interpretation.} All datasets maintain significant faithfulness under region+composition matching (all $d_z > 0.5$, all win rates $> 70\%$), confirming that saliency captures position-specific patterns beyond regional sensitivity. The reduced effect sizes (approximately 15--20\% lower $d_z$) indicate that some portion of the composition-only effect reflects regional clustering, but the majority of the signal is position-specific. This robustness check addresses the main confound critique.

\section{Empirical Validation Metrics}

\subsection{BioPrior Effectiveness}
\begin{itemize}
    \item \textbf{Constraint satisfaction rate:} Percentage of designed sequences satisfying all constraints
    \item \textbf{Loss reduction:} $\Delta\mathcal{L}_{\text{bio}} = \mathcal{L}_{\text{bio}}^{\text{initial}} - \mathcal{L}_{\text{bio}}^{\text{final}}$
    \item \textbf{Gradient norm:} $\|\nabla\mathcal{L}_{\text{bio}}\|$ throughout training
\end{itemize}

\subsection{Perturbation Test Metrics}
\begin{itemize}
    \item \textbf{Wilcoxon p-value:} Significance of $\Delta(T) > \Delta(R)$
    \item \textbf{Cohen's $d_z$:} Standardized effect size
    \item \textbf{Win rate:} Proportion of samples with $\Delta(T) > \Delta(R)$
    \item \textbf{Composition bias:} $B = \mathbb{E}[\Delta_{\text{unmatch}}] - \mathbb{E}[\Delta_{\text{match}}]$
\end{itemize}

\section{Model Implementation Details}
\label{sec:implementation}

\subsection{Architecture}
Our model uses separate siRNA and mRNA encoders followed by bi-directional cross-attention. Each encoder applies a shallow convolutional front-end, a 2-layer BiLSTM ($\texttt{hidden}=32$ per direction), and a Transformer encoder with $H=8$ heads and $n_{\text{layers}}=1$ on top of the BiLSTM outputs. We then apply cross-attention in both directions (siRNA$\rightarrow$mRNA and mRNA$\rightarrow$siRNA), pool each stream by concatenating mean and max pooling, and concatenate pooled FM features and thermodynamic features $\mathbf{t}_d$ before the final MLP classifier. Dropout is 0.12 in the encoder stack and 0.15 in the prediction MLP.

\subsection{Biology-Derived Input Features}
The siRNA input is enhanced with biological features computed from the one-hot encoding: position importance (uniform), seed region indicator (positions 2-8, 1-indexed), cleavage site indicator (positions 9-11), global GC content, seed AU content, thermodynamic asymmetry ($\mathrm{GC}_{3'} - \mathrm{GC}_{5'}$), and per-position binary indicators for AU and GC nucleotides.

\subsection{Training Objective}
We train with the combined objective:
\[
\mathcal{L}_{\text{total}} = \mathcal{L}_{\text{pred}} + \lambda(t)\mathcal{L}_{\text{bio}} + \lambda_{\text{aux}}\mathcal{L}_{\text{aux}}
\]

\textbf{Regression-to-classification evaluation:} The model outputs a continuous efficacy score $\hat{y} \in [0,1]$ and is trained with weighted MSE loss against normalized efficacy values. For classification metrics (ROC-AUC, PR-AUC, F1), we binarize ground-truth labels at threshold $\tau = 0.7$ (``high efficacy''). F1-optimal thresholds are selected on validation folds by sweeping $[0.3, 0.75]$; this is standard practice but means reported F1 values are optimistic relative to held-out threshold selection. ROC-AUC and PR-AUC are threshold-free ranking metrics.

We upweight high-efficacy samples ($y\ge 0.7$) to mitigate label imbalance:
\[
w_i = \begin{cases}
2.0 & y_i \ge 0.7 \\
1.0 & \text{otherwise}
\end{cases}
\]

\subsection{Optimization}
We use AdamW \citep{loshchilov2019adamw} with learning rate $10^{-4}$ and weight decay $10^{-4}$, batch size 32. We use linear warmup for 5 epochs followed by cosine annealing with warm restarts every 100 epochs; early stopping is based on validation ROC-AUC (patience 30). Unless stated otherwise, models are trained up to 300 epochs.

\textbf{Sample weighting note:} We precompute sample weights based on efficacy labels and align them with batch order. Data loading uses \texttt{shuffle=False} to maintain this alignment; for distributed or shuffled training, weights should be included in the dataset's \texttt{\_\_getitem\_\_} return value.

\subsection{Cross-validation and Splits}
We perform stratified 5-fold cross-validation within each dataset using the high-efficacy threshold $y\ge 0.7$ for stratification. In each fold, one fold is held out for testing; from the remaining folds we reserve 20\% as validation for early stopping and model selection. We report mean and standard deviation across folds.

\subsection{Hyperparameters}
\begin{table}[h]
\centering
\begin{tabular}{lcc}
\toprule
\textbf{Parameter} & \textbf{Value} & \textbf{Description} \\
\midrule
siRNA length & 19 & Nucleotide positions \\
Batch size & 32 & Training batch size \\
Learning rate & $10^{-4}$ & AdamW initial learning rate \\
Weight decay & $10^{-4}$ & L2 regularization \\
Dropout (encoder) & 0.12 & Encoder dropout rate \\
Dropout (MLP) & 0.15 & Classifier dropout rate \\
BiLSTM hidden & 32 & Hidden size per direction \\
Transformer heads & 8 & Multi-head attention heads \\
Transformer layers & 1 & Transformer encoder layers \\
\bottomrule
\end{tabular}
\caption{Key hyperparameters}
\label{tab:hyperparams}
\end{table}

\subsection{Compute Environment}
Experiments were run on NVIDIA A100 40GB GPUs. Typical training time per fold: 2-3 hours for 300 epochs. Perturbation testing for 200 samples requires approximately 30 minutes. We use PyTorch 1.12.0, CUDA 11.3, Python 3.9.

\subsection{Reproducibility}
\textbf{Random seeds:} All experiments use \texttt{torch.manual\_seed(42)} and \texttt{torch.cuda.manual\_seed\_all(42)} at the start of each fold. KFold splits use \texttt{random\_state=42}. NumPy random state is set via \texttt{np.random.seed(42)}.

\textbf{Determinism:} We set \texttt{torch.backends.cudnn.deterministic=True} and \texttt{torch.backends.cudnn.benchmark=False} for reproducible results, though this incurs a $\sim$10\% training slowdown.

\textbf{Code and data:} Code will be released upon publication. Datasets are publicly available: Hu~\citep{huesken2005design}, Mix~\citep{reynolds2004rational}, Taka~\citep{katoh2007sirna}, Shabalina~\citep{shabalina2006computational}.

\section{Full Inter-Dataset Transfer Results}
\label{app:transfer_faith}
\label{app:full_results}

Table~\ref{tab:transfer_full} presents complete inter-dataset transfer results with saliency faithfulness validation for all 12 source-target pairs.

\begin{table}[h]
\centering
\small
\begin{tabular}{llccccc}
\toprule
\textbf{Source} & \textbf{Target} & \textbf{AUC} & \textbf{PCC} & \textbf{Win \%} & \textbf{$d_z$} & \textbf{Status} \\
\midrule
Hu & Mix & 0.773 & 0.54 & 98.5 & 1.54 & \cmark \\
Hu & Taka & 0.535 & 0.06 & 100 & 1.54 & \cmark \\
Hu & Shabalina & 0.698 & 0.47 & 100 & 1.87 & \cmark \\
\midrule
Mix & Hu & 0.792 & 0.57 & 92.0 & 1.23 & \cmark \\
Mix & Taka & 0.497 & 0.01 & 97.5 & 1.47 & \cmark \\
Mix & Shabalina & 0.712 & 0.48 & 88.5 & 0.59 & \cmark \\
\midrule
Shabalina & Hu & 0.787 & 0.55 & 70.0 & 0.45 & \cmark \\
Shabalina & Mix & 0.816 & 0.63 & 75.5 & 0.56 & \cmark \\
Shabalina & Taka & 0.559 & 0.13 & 61.0 & 0.35 & \cmark \\
\midrule
Taka & Hu & 0.490 & -0.01 & 9.5 & -1.25 & \xmark \\
Taka & Mix & 0.510 & -0.01 & 7.6 & -1.37 & \xmark \\
Taka & Shabalina & 0.517 & 0.04 & 9.5 & -1.30 & \xmark \\
\bottomrule
\end{tabular}
\caption{Complete inter-dataset transfer results (+BioPrior model). Models trained on Hu/Mix/Shabalina maintain faithful saliency (9/9 pass) regardless of prediction performance. Models trained on Taka exhibit inverted saliency on all other datasets (0/3 pass), with high-saliency positions being less important than random. Win rates near 100\% occur when the matched baseline is tightly concentrated; median per-sample differences $d_j = \Delta(T_j) - \Delta_{\text{match},j}$ confirm genuine separation (Hu$\to$Taka: median $d_j = 0.08$, IQR $[0.04, 0.13]$).}
\label{tab:transfer_full}
\end{table}

\paragraph{Interpretation.} The 9/9 pass rate for non-Taka sources versus 0/3 for Taka sources confirms that the failure is source-specific rather than target-specific. Taka-trained models learn position-importance patterns (primary peak at 9--11, secondary at 16--19) that are inversely related to efficacy determinants in other datasets (5$'$ terminus dominance at positions 1--4), resulting in inverted saliency when transferred.

\paragraph{Why separate datasets rather than pooling?}
A natural question is why we evaluate on four separate datasets rather than combining them into a single larger training set. We deliberately maintain dataset separation for three reasons:

First, \textbf{pooling obscures protocol-specific effects}. Our transfer experiments reveal that Taka exhibits fundamentally different position-importance patterns (primary peak at 9--11, secondary at 16--19) compared to Hu, Mix, and Shabalina (5$'$ terminus dominance at positions 1--4). Pooling would mask this heterogeneity, producing a model that compromises between incompatible biological signals without revealing the underlying conflict.

Second, \textbf{separate evaluation enables diagnosis}. By training on each dataset independently and measuring cross-dataset transfer, we can identify which dataset pairs share compatible biological patterns and which do not. This diagnostic capability would be lost with pooled training, where failures would manifest as reduced overall performance without clear attribution.

Third, \textbf{real-world deployment requires protocol awareness}. Practitioners developing siRNA therapeutics typically work within a specific experimental protocol. Our results demonstrate that a model validated on one protocol (e.g., Hu's branched DNA assay) may not generalize to another (e.g., Taka's luciferase reporter). Separate evaluation provides actionable guidance: models transfer reliably among Hu/Mix/Shabalina but should not be applied to Taka-like protocols without retraining.

Finally, \textbf{pooling does not guarantee improved performance}. Preliminary experiments with combined Hu+Mix+Shabalina training showed marginal gains on held-out samples from these datasets but no improvement on Taka transfer, confirming that the Taka incompatibility is fundamental rather than a sample size limitation.

\subsection{Intra-Dataset Fold-Level Analysis}
\label{app:fold_analysis}

\begin{figure}[h]
\centering
\includegraphics[width=\textwidth]{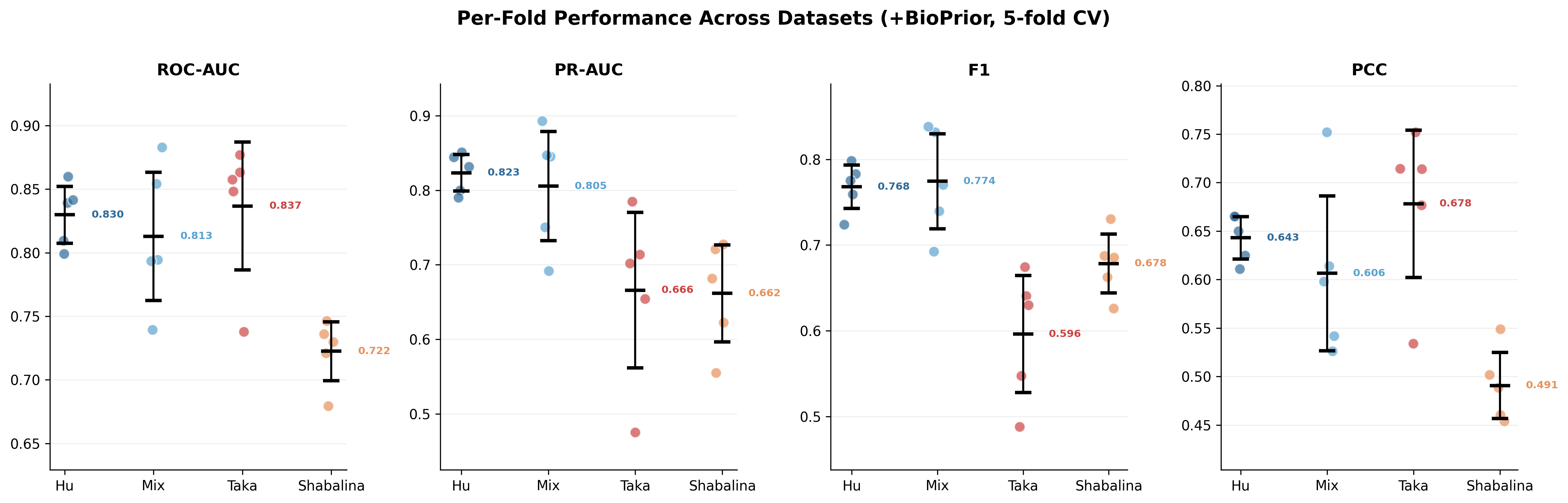}
\caption{\textbf{Per-fold performance across datasets (+BioPrior, 5-fold CV).} Individual fold results (circles) with mean $\pm$ std (black bars). Taka shows the largest inter-fold variance, particularly for PR-AUC (fold~3: 0.475 vs.\ fold~1: 0.785). Mix fold~1 is a low-AUC outlier (0.739) compared to the other four folds ($>$0.79). Shabalina has the most consistent fold-level performance (CV $<$ 4\% for AUC).}
\label{fig:fold_strip}
\end{figure}

\begin{figure}[h]
\centering
\includegraphics[width=0.65\textwidth]{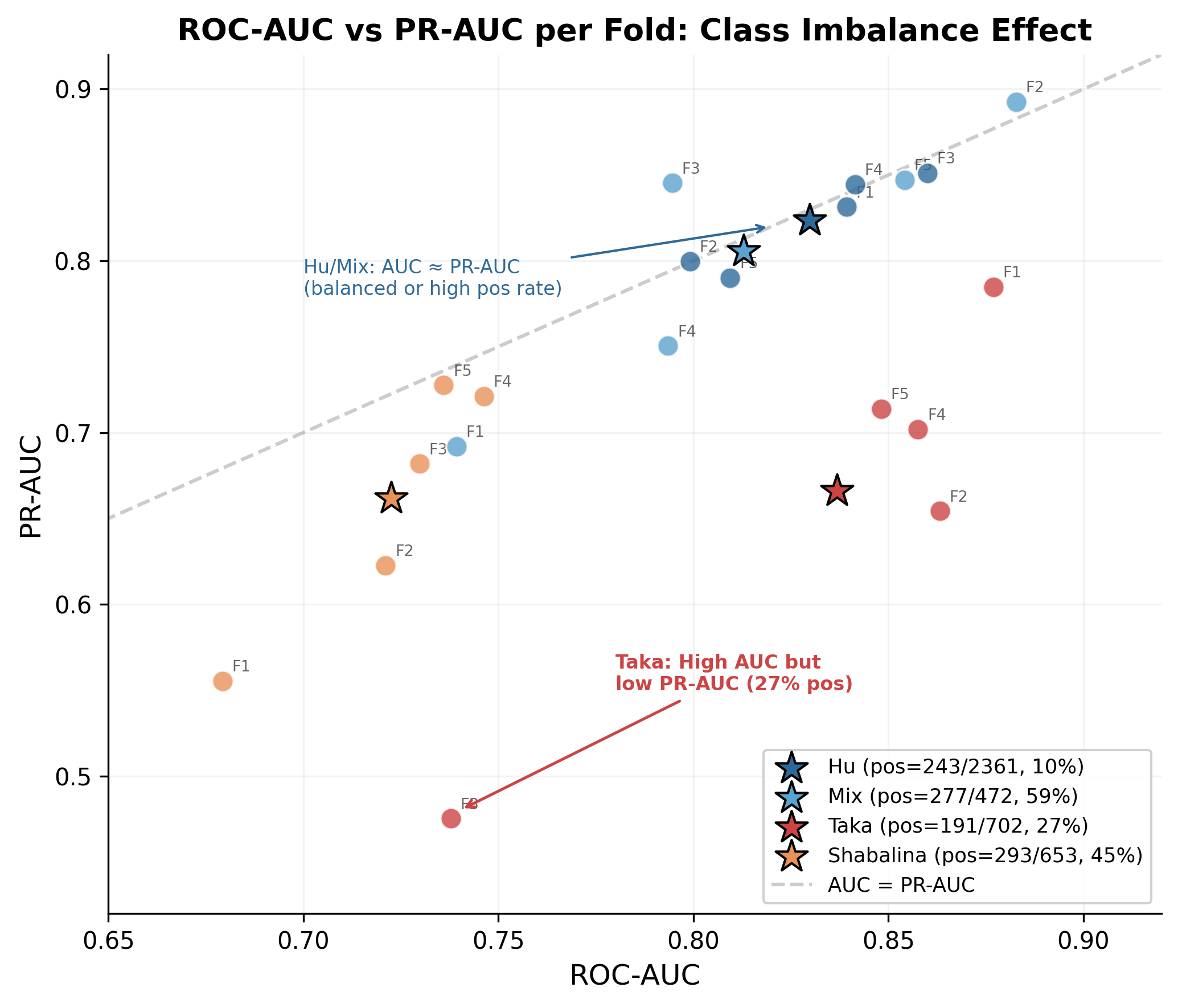}
\caption{\textbf{ROC-AUC vs.\ PR-AUC per fold, revealing class imbalance effects.} Each small dot is one fold; stars mark dataset means. Points below the diagonal indicate that PR-AUC $<$ AUC, which occurs when the positive class is rare. Hu (10\% positive rate) and Taka (27\%) exhibit the largest AUC--PR-AUC gaps. Notably, Taka achieves competitive AUC (0.84 mean) but substantially lower PR-AUC (0.67 mean), indicating that ROC-AUC alone overestimates ranking quality for imbalanced datasets, motivating our inclusion of PR-AUC in Table~\ref{tab:predictive}.}
\label{fig:auc_prauc_scatter}
\end{figure}

Figures~\ref{fig:fold_strip} and~\ref{fig:auc_prauc_scatter} provide per-fold granularity for the summary statistics reported in Table~\ref{tab:predictive}.

\subsection{Transfer Ablation: Baseline vs.\ BioPrior}
\label{app:transfer_ablation}

\begin{figure}[h]
\centering
\includegraphics[width=\textwidth]{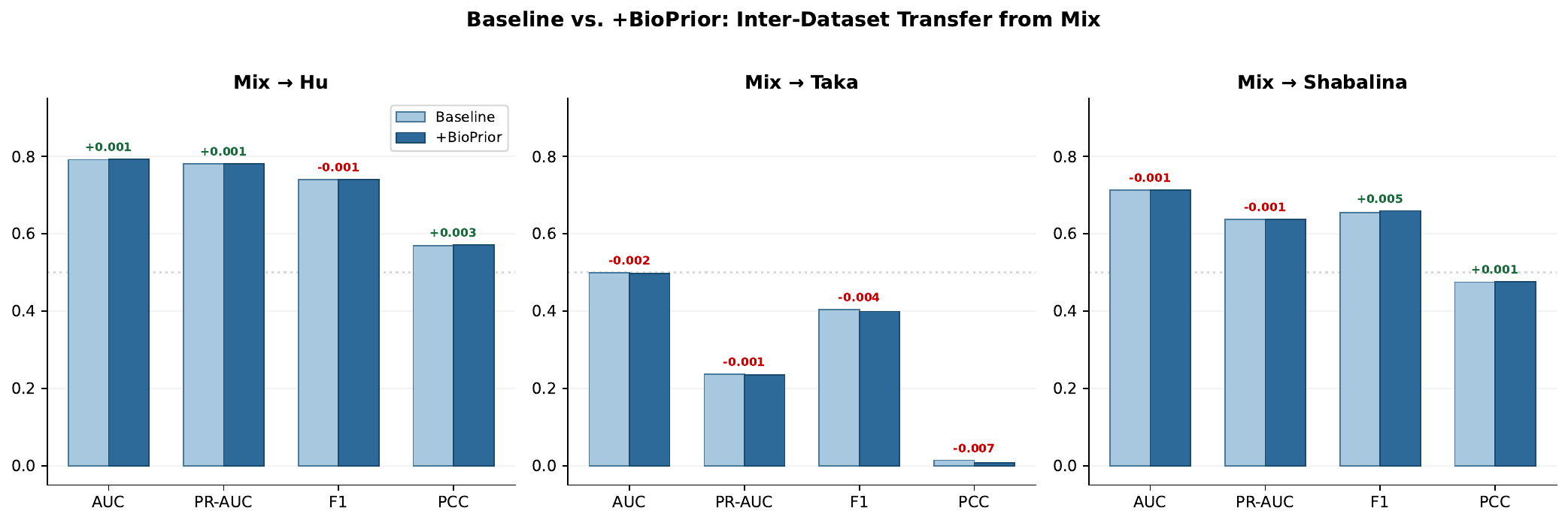}
\caption{\textbf{Baseline vs.\ +BioPrior on Mix-sourced inter-dataset transfers.} For each target dataset, we compare AUC, PR-AUC, F1, and PCC between the baseline model and the BioPrior-regularized model. BioPrior produces marginal improvements on Mix$\to$Hu (+0.001 AUC, +0.001 PR-AUC) and Mix$\to$Shabalina (+0.005 F1), but provides no benefit on Mix$\to$Taka (both models achieve $\approx$0.50 AUC). This confirms that biology-informed constraints cannot rescue fundamentally misaligned dataset pairs.}
\label{fig:transfer_ablation}
\end{figure}

\end{document}